\documentclass[twocolumn]{aastex61}

\newcommand\aastex{AAS\TeX}

\received{23 January 2018}
\revised{27 February 2018}
\accepted{3 March 2018}

\submitjournal{ApJ}

\shorttitle{\aastex\ A Near-infrared RR~Lyrae census along the southern Galactic plane}
\shortauthors{D\'ek\'any et al.}

\begin{document}

\title{A Near-infrared RR~Lyrae census along the southern Galactic plane: the Milky Way's stellar fossil brought to light}
\correspondingauthor{Istv\'an D\'ek\'any}
\email{dekany@uni-heidelberg.de}

\author{Istv\'an D\'ek\'any}
\affiliation{Astronomisches Rechen-Institut, Zentrum f\"ur Astronomie der Universit\"at Heidelberg,
M\"onchhofstr. 12-14, 69120 Heidelberg, Germany}

\author{Gergely Hajdu}
\affiliation{Astronomisches Rechen-Institut, Zentrum f\"ur Astronomie der Universit\"at Heidelberg,
M\"onchhofstr. 12-14, 69120 Heidelberg, Germany}
\affiliation{Instituto de Astrof\'isica, Facultad de F\'isica, Pontificia Universidad Cat\'olica de Chile, Av. Vicu\~na Mackenna 4860, 782-0436 Macul, Santiago, Chile}
\affiliation{Millennium Institute of Astrophysics, Santiago, Chile}

\author{Eva K. Grebel}
\affiliation{Astronomisches Rechen-Institut, Zentrum f\"ur Astronomie der Universit\"at Heidelberg,
M\"onchhofstr. 12-14, 69120 Heidelberg, Germany}

\author{M\'arcio Catelan}
\altaffiliation{On sabbatical leave at the Astronomisches Rechen-Institut, Zentrum f\"ur Astronomie der Universit\"at Heidelberg, Heidelberg, Germany.}
\affiliation{Instituto de Astrof\'isica, Facultad de F\'isica, Pontificia Universidad Cat\'olica de Chile, Av. Vicu\~na Mackenna 4860, 782-0436 Macul, Santiago, Chile}
\affiliation{Millennium Institute of Astrophysics, Santiago, Chile}

\author{Felipe Elorrieta}
\affiliation{Departmento de Estad\'istica, Facultad de Matem\'aticas, Pontificia Universidad Cat\'olica de Chile, Av. Vicu\~na Mackenna 4860, 7820436 Macul, Santiago, Chile}
\affiliation{Millennium Institute of Astrophysics, Santiago, Chile}

\author{Susana Eyheramendy}
\affiliation{Departmento de Estad\'istica, Facultad de Matem\'aticas, Pontificia Universidad Cat\'olica de Chile, Av. Vicu\~na Mackenna 4860, 7820436 Macul, Santiago, Chile}
\affiliation{Millennium Institute of Astrophysics, Santiago, Chile}
\affiliation{Max-Planck-Institut f\"ur Astronomie, K\"onigstughl 17, 69117 Heidelberg, Germany}

\author{Daniel Majaess}
\affiliation{Mount Saint Vincent University, Halifax, Nova Scotia, Canada}
\affiliation{Saint Mary's University, Halifax, Nova Scotia, Canada}

\author{Andr\'es Jord\'an}
\affiliation{Instituto de Astrof\'isica, Facultad de F\'isica, Pontificia Universidad Cat\'olica de Chile, Av. Vicu\~na Mackenna 4860, 782-0436 Macul, Santiago, Chile}
\affiliation{Millennium Institute of Astrophysics, Santiago, Chile}
\affiliation{Max-Planck-Institut f\"ur Astronomie, K\"onigstughl 17, 69117 Heidelberg, Germany}

\begin{abstract}

RR~Lyrae stars (RRLs) are tracers of the Milky Way's fossil record, holding valuable information on its formation and early evolution. Owing to the high interstellar extinction endemic to the Galactic plane, distant RRLs lying at low Galactic latitudes have been elusive. We attained a census of 1892 high-confidence RRLs by exploiting the near-infrared photometric database of the VVV survey's disk footprint spanning $\sim$70$^\circ$ of Galactic longitude, using a machine-learned classifier. Novel data-driven methods were employed to accurately characterize their spatial distribution using sparsely sampled multi-band photometry. The RRL metallicity distribution function (MDF) was derived from their $K_s$-band light curve parameters using machine-learning methods. The MDF shows remarkable structural similarities to both the spectroscopic MDF of red clump giants and the MDF of bulge RRLs. We model the MDF with a multi-component density distribution and find that the number density of stars associated with the different model components systematically changes with both the Galactocentric radius and vertical distance from the Galactic plane, equivalent to weak metallicity gradients. Based on the consistency with results from the ARGOS survey, three MDF modes are attributed to the old disk populations, while the most metal-poor RRLs are probably halo interlopers. We propose that the dominant [Fe/H] component with a mean of $-1$\,dex might correspond to the outskirts of an ancient Galactic spheroid or classical bulge component residing in the central Milky Way. The physical origins of the RRLs in this study need to be verified by kinematical information.

\end{abstract}


\section{Introduction} \label{sec:intro}

RR Lyrae (RRL) stars are keystone objects of Galactic archeology. They are truly stellar fossils -- being over 10 Gyr old, they already existed when the primary structures of the Milky Way formed, thus they are excellent proxies of its primordial stellar populations and carry precious information on the early formation history of our Galaxy \citep[e.g.,][]{2015pust.book.....C}. 
They follow precise period-luminosity-metallicity relations (PLZRs) in the infrared (IR) wavebands \citep[e.g.,][and references therein]{2004ApJ...610..269D,2015ApJ...799..165B,2017A&A...604A.120N}, enabling us to employ them as primary standard candles. Moreover, they can be employed as relatively precise photometric metallicity indicators in two different ways. First, the shapes of their light curves are tightly related to their heavy element abundances \citep{1996A&A...312..111J,2005AcA....55...59S}; and second, by inverting their near-IR PLZRs, the metallicities of RRL populations of known distances can be determined with high precision \citep{2016MNRAS.461L..41M,2016AJ....152..170B}.

Since they are fairly bright objects ($M_K\simeq0.5$), their census has been feasible in all Galactic components, as well as in the surrounding dwarf satellites and stellar streams \citep[e.g.,][]{2013ApJ...763...32D,2013AJ....146...21S,2014AcA....64..177S,2016AcA....66..131S,2015MNRAS.454.1509M,2017A&A...599A.125F}. However, a deep survey of RRL stars throughout the Galactic disk has been lacking due to the severe interstellar extinction endemic to objects lying close to the Galactic plane, thus limiting the RRL census to within a few kiloparsecs of the solar neighborhood along low-latitude sightlines \citep[e.g.,][]{1994AJ....108.1016L}.

Owing to their exceptional diagnostic value, RRL stars along the Galactic disk hold the potential to add key pieces of information to our current understanding of the structural and chemical evolution of stellar populations in the disk, bulge, and halo, and the interplay between them. 
In recent years, there has been enormous advancement in the characterization of the metallicity distribution function (MDF) throughout the bulge and the Galactic disk, which was made possible by the datasets from large spectroscopic surveys \citep[e.g.,][]{2013MNRAS.430..836N,2014ApJ...796...38N,2015ApJ...808..132H}. 
While these data serve as a solid foundation for studies of Galaxy evolution by providing precise abundances for a very large number of stars, they are currently unable to put strong constraints on stellar ages and distances, although the latter issue is well mitigated by the sheer quantity of data. 
Consequently, the resulting stellar parameter space sampled by current spectroscopic surveys represents a blend of stellar populations with a largely unknown age distribution, which complicates the interpretation of the observations. 
Complementary datasets from tracer objects with significant constraints on their ages and positions are therefore key pieces of the puzzle, even though their sample sizes can be 1--2 orders of magnitude smaller. 
In this context, the census of RRL stars in the Galactic disk is of the utmost importance because they provide a unique access to the oldest fossil record of the Milky Way.

In a recent pilot study \citep{2017AJ....153..179M}, we leveraged near-IR data from the VISTA Variables in the V\'ia L\'actea survey \citep[VVV,][]{2010NewA...15..433M} to identify fundamental-mode RRL (RRab) stars lying in close proximity of the southern Galactic plane. 
We presented a sample of 404 candidate RRab stars in a narrow strip of the sky at $-2.24^\circ\lesssim b \lesssim -1.05^\circ$ running parallel with the Galactic equator between $295^\circ \lesssim l \lesssim 350^\circ$, covering $25\%$ of the VVV disk footprint. 
Based on their $J,H,K_s$ photometry, we performed a preliminary characterization of the objects. 
The present study expands upon this earlier work in several aspects. 
We extend our census to the entire area of the southern disk footprint of the VVV survey, and apply novel methods for the identification and accurate characterization of the RRL stars based on near-IR photometry alone. 

This paper is organized as follows. In Section~\ref{sec:search}, we describe the steps of data processing and the identification of RRab stars by means of a machine-learned classifier. 
Section~\ref{sec:charact} discusses the methodology for the characterization of the RRab sample based on near-IR photometry, including the measurement of accurate mean magnitudes in the $JHK_s$ bands, estimating the extinctions of the stars, the estimation of their iron abundances from their $K_s$-band light curves, and the computation of the distances to the objects. 
In Section~\ref{sec:map}, we present their spatial and metallicity distributions inferred from their light curves. 
We compare our results to former studies and discuss their implications for the early formation history of the Milky Way in Section~\ref{sec:discussion}. We summarize our conclusions in Section~\ref{sec:summary}.

\section{The census of disk RR Lyrae stars} \label{sec:search}

\subsection{Data} \label{subsec:data}

Our study is entirely based on near-IR time-series photometric data from the VVV survey. 
We analyzed observational data from the VVV's disk footprint (fields d001--d151, as defined in \citealt{2010NewA...15..433M}), an approximately $4^{\circ}\times57^{\circ}$ elongated area running parallel with and nearly symmetrically to the Galactic equator, covering the mid-plane toward the 4th Galactic quadrant. The data were acquired between January 30, 2010 and August 14, 2015.
Images were taken in the $ZYJHK_s$ broadband filter set of the VISTA photometric system. Photometric zero-points (ZPs) were calibrated using local secondary standard stars from the 2MASS survey \citep{2006AJ....131.1163S}. 
Each field was observed 50 times in the $K_s$ band, and a few (1--10) times in the $Z,Y,J$ and $H$ bands. Since the tiles slightly overlap, a small fraction of the survey area was observed in up to 200 epochs. 

The analysis presented in the following sections is based on the standard, public VVV data products provided by the Cambridge Astronomy Survey Unit (CASU) and available through the ESO Archive. 
Observational data were pipeline-processed by the VISTA Data Flow System \citep[VDFS; see][]{2004SPIE.5493..401E}. We used ZP-calibrated magnitudes measured in a series of small, flux-corrected circular apertures. 
The VDFS is shown to provide accurate magnitudes in moderately crowded fields \citep{2004SPIE.5493..411I}. 
Limiting magnitudes range from $\sim$18.5 to $\sim$17~mag in the $K_s$ band, and between $\sim$18 and $\sim$20~mag in the $J$ band, mainly depending on the interstellar extinction and source crowding along the line of sight \citep[see also][]{2012A&A...537A.107S}.

Positional cross-matching of the photometric source tables provided by CASU was performed using software code from the Starlink Tables Infrastructure Library \citep[{\sc STIL},][]{2006ASPC..351..666T} on a tile-by-tile basis. 
First, fiducial source lists were computed from individual distortion-corrected source catalogs extracted from contiguous mosaic images ({\em tiles}), generated by the VDFS from six non-contiguous detector frame stacks ({\em pawprints}), which were acquired sequentially at each observational epoch by offset pointing. 
For each field, the 20 highest-quality tile-based source catalogs were identified in terms of best seeing, smallest mean source ellipticity, and the lowest atmospheric foreground flux. 
These source lists were randomly distributed into groups of five, and were reduced into single source lists via the multi-object positional cross-matching of the objects' celestial positions with a tolerance of $0.36''$ (=1 detector pixel), and keeping the mean positions of the best `friends of friends' matching groups. 
The same procedure was then repeated on the resulting four merged catalogs, producing a single master source list. We note that this approach is robust against spurious sources and systematic positional mismatches in crowded areas due to changing seeing conditions, but has the drawback of missing transient events and very faint objects that only occasionally emerge from the background, and cannot identify objects with high proper motion. 
But since our goal is to identify distant RRLs with sufficiently high-quality light curves enabling their firm classification, these drawbacks are of no significance in the present study.

The master source catalogs contain $6\cdot10^5$---$10^6$ point sources per field, depending on overall source density and extinction. 
Photometric time-series were obtained from pawprint-based VDFS source catalogs by performing simple one-by-one pairwise positional cross-matching between the celestial positions of the objects in these tables and those in the master source catalog, then sorting the union of the resulting tables by object identifier and Julian date. 
We note that although tile- and pawprint-based photometries are computed by the same VDFS procedures, we opted to use the latter because, due to fewer image processing steps, they contain less photometric systematics at the expense of slightly brighter limiting magnitudes.

\subsection{Variability search} \label{subsec:var}

The post-processing of the VDFS data products discussed in Section~\ref{subsec:data} resulted in approximately $2\cdot10^8$ photometric time-series from the VVV disk area. 
To find RRLs in this vast amount of data, we first pre-selected a subset of objects showing putative light variations by taking advantage of the correlated sampling of the light curves. 
Due to the observing strategy of VISTA, each field is covered by 6 consecutive exposures at each epoch within a $\sim$3-minute interval, in order to obtain offset images for covering the gaps between the detector's 16 chips. 
Since we use pawprint-based photometry, in the overwhelming majority of the light curves, the measurements are clustered into 2--6 points per epoch (depending on the position of the object on the detector), which originate from the same tile acquisition sequence, and have a time span that is negligible compared to the time-scale of the light variation of RRLs.

We utilize this property of the data to search for objects with coherent light variations by employing two different variability statistics designed for correlated sampling, namely Stetson's $J$ index \citep[][Eq.~1]{1996PASP..108..851S}, and the ratio of the weighted standard deviation of the time-series $\sigma_w$ and a pseudo von Neumann index \citep{1941AMS..12..153} $\delta$. 
In the computation of $\delta$, the standard form of mean squared successive differences (which measure the point-to-point scatter in the data) was modified by including the squared inverses of the measurement errors as weights, and by evaluating the summations for only those successive measurement pairs that belong to the same tile acquisition sequence, i.e.,
\begin{equation}
\delta^2=\frac{\sum_{i=1}^{N-1}(m_{i+1}w_{i+1}-m_iw_i)^2}{\frac{M-1}{M}\sum_{i=1}^{N}w_i},
\end{equation}

\noindent where $N$ is the number of appropriate measurements, $M$ is the number of nonzero weights (in practice, $N=M$), and $w_i=\sigma_i^{-2}$, where $\sigma_i^{2}$ is the squared sum of the photometric and ZP errors of magnitude $m_i$.

Candidate variable stars were selected above the $0.1\%$ significance level of either $J$ or $\sigma_w/\delta$, as estimated from Monte Carlo simulations based on Gaussian noise. 
We note that this first statistical filter is not effective against strong photometric systematics (a.k.a `colored noise'), but still reduces the objects of interest to a few percent of all point sources.

In order to mitigate the flux contamination due to source crowding and maximize the photometric signal-to-noise ratio ($S/N$), it is necessary to determine the optimal aperture for each star. In the first stage of the analysis, the selection of the optimal aperture at first approximation was done by minimizing the global scatter of the magnitudes, in conjunction with a basic iterative $5\sigma$ outlier rejection and various parameter-threshold rejections using metadata computed by CASU (i.e., photometric confidence, source ellipticity, etc.).
Based on the variability statistics, typically $\sim10^4$ objects passed this first candidate selection in each VVV field, and were then propagated into a period analysis.

\subsection{Period search and light curve model} \label{subsec:per}

We searched for periodic signals in the $K_s$-band time-series using our own parallelized implementation of the Generalized Lomb--Scargle periodogram \citep[GLS,][]{2009A&A...496..577Z}. 
For this purpose, individual measurements were binned to form light curves with one point per epoch, in order to provide a consistent amplitude scale in frequency space. 
Spectral significance levels were estimated analytically (see \citealt{2009A&A...496..577Z} for details), and stars with periodic signals in the $[0.35\,{\rm day},0.9\,{\rm day}]$ range with a false alarm probability (FAP) of $<0.1\%$ were selected for further analysis. 
Bootstrap simulations with a subset of data showed that the analytical approximation of the FAP always underestimates the significance for our data, thus our sample contains more false positives compared to a selection based on a star-by-star bootstrap-based estimation of the FAP. 
However, the latter would inflict a dramatic increase in the computational cost, and would only increase sample purity for low $S/N$ observations. 
Since we are interested in RRLs with light curves of sufficiently high quality (allowing accurate classification), our simple approach of FAP estimation does not affect the final sample completeness and purity, but merely increases the sample size in the later classification analysis by the inclusion of noisy data, still resulting in an enormous overall gain in the computational cost.

Our period analysis resulted in a reduced {\em target sample} of approximately $85000$ candidate objects, which were subjected to an iterative feature extraction procedure. 
In the first iteration, a light curve was fitted in each aperture with a truncated Fourier series model using the GLS period, and the optimal aperture was reevaluated by minimizing the reduced $\chi^2$.
The number of Fourier terms was optimized following a similar procedure as described by \cite{2007A&A...462.1007K}. 
In each subsequent iteration, the period was refitted, followed by a $3\sigma$ outlier rejection around the updated model. 
The parameters of the final model solution resulting from this procedure were used as input features for classification.

\subsection{Light curve classification} \label{subsec:classif}

In the next step of the analysis, we employed a machine-learned classifier developed by \cite{2016A&A...595A..82E} for identifying fundamental-mode RRL (RRab) stars in our database. 
The procedure uses the adaptive boosting algorithm to assign a score to each light curve, based on a set of 12 features, which are parameters of the light curve model and descriptive statistics of the time-series (see \citealt{2016A&A...595A..82E} for a full description of the features). 
The classifier was trained using data from 3 fields of the VVV survey's bulge area (namely b293, b294, b295) overlapping with the footprint of the OGLE-IV survey, where the census of RRab stars is close to complete. 
The training set consisted of several hundred RRab stars (with firm classification based on optical light curves), as well as tens of thousands of non-RRab light curves from the same fields.
The decision boundary (i.e., a score threshold) that optimizes the classifier performance was computed by tenfold cross-validation \citep[for a full explanation, see][]{2016A&A...595A..82E}.
In our original analysis \citep{2016A&A...595A..82E}, we set the decision boundary at a score threshold of 0.548, by optimizing the classifier based on the so-called $F_1$-measure, which is the harmonic mean between precision and recall (in other terms, sample purity and completeness). 
This threshold corresponded to a precision of $\sim$0.97, and a recall of $\sim$0.90 \citep{2016A&A...595A..82E}.

Since the observing strategy of the VVV survey was slightly different for its bulge and disk areas, we found it desirable to complement earlier measures of classifier performance (which were based on data from the bulge area) with a performance estimate based on data from the disk area (which our present study focuses on). 
The reason for this is that the photometric time-series from the disk area have a smaller number of epochs ($\sim$50) compared to the bulge data ($\sim100$), as well as a slightly smaller baseline. 
Since the accuracy of the features used by the classifier depend on the distribution of data points \citep[see][]{2016A&A...595A..82E}, these differences in the data properties are expected to result in a slightly lower classifier performance for the disk area at any given score threshold compared to our original estimates. 
At the same time, we are aiming to study an area with practically no previous census of RRab stars, thus we cannot compare our class predictions to already labeled data (i.e., RRab stars with accurate classifications based on independent data). 
We note that the OGLE-III survey identified RRab stars in a small number of fields along the southern Galactic disk, but there is only a very small area in overlap with the VVV fields, containing only 6 previously known RRab stars.

\begin{figure}[]
\includegraphics[width=0.45\textwidth]{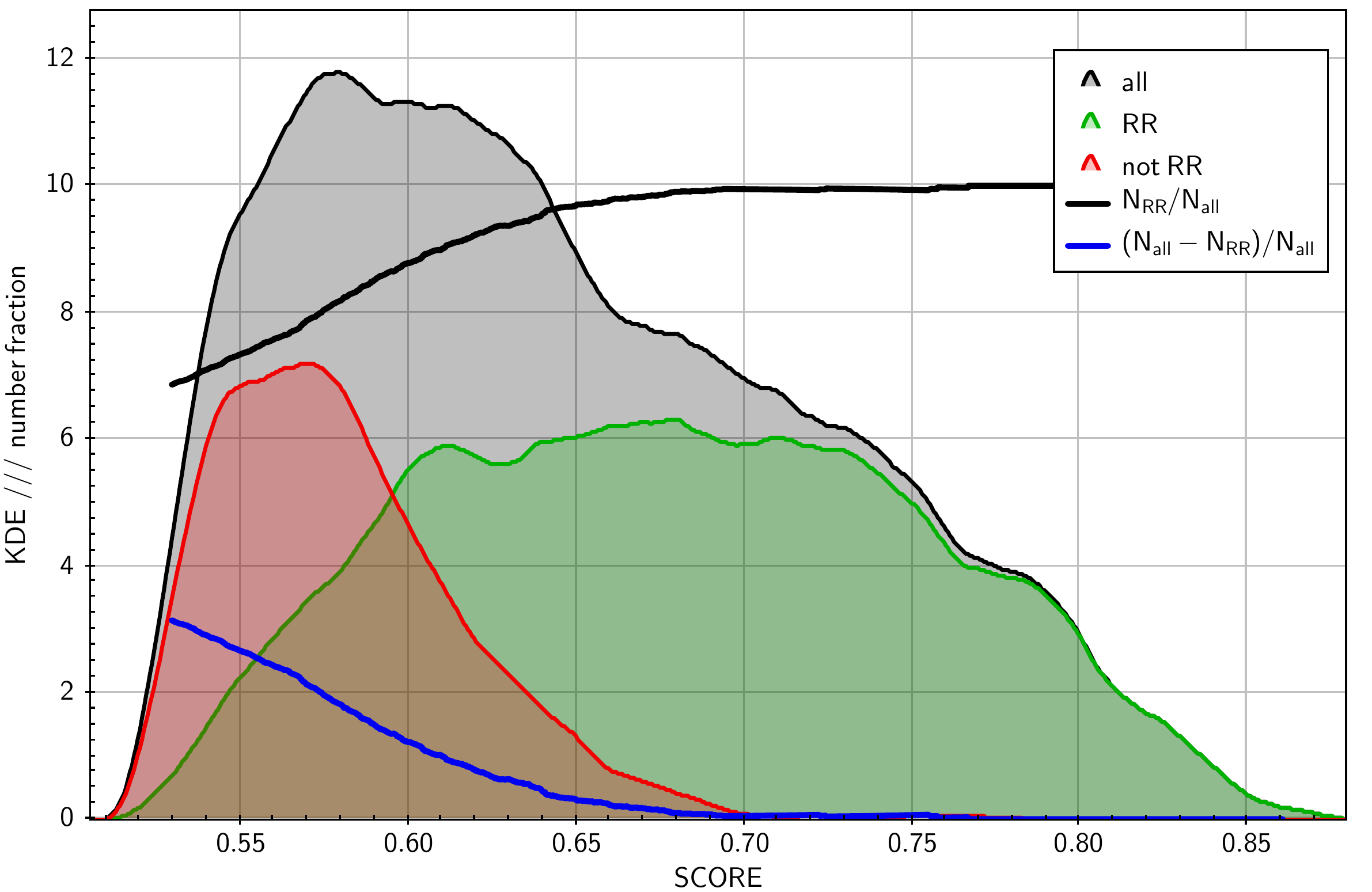}
\caption{Kernel density estimates (KDE) of the true and false positives and their sum (shaded green, red, and gray areas, respectively) as a function of classifier score for our {\em broad selection} of RRab candidates, resulting from visual inspection by domain experts. Black and blue curves show the corresponding precision and contamination, respectively, multiplied by 10 ({\em see the text}).\label{fig:precision}}
\end{figure}

In order to estimate the precision of the classifier on VVV disk data, first we ran the algorithm on our entire {\em target sample} (see Sect.~\ref{subsec:per}), and applied a score threshold of 0.5, resulting in a {\em broad selection} of 3379 objects. 
Then we divided this into groups of 500 objects each by random selection. 
Each of these groups were visually inspected by human domain experts, and labeled as RRab / non-RRab, {\em without knowledge of the score}, in order not to be biased by the result from the classifier. 
We note that not every human expert inspected all light curves in the broad selection, but all light curves were inspected by at least one human. 
Based on those objects that were inspected by multiple humans, we concluded that the results from different humans had a consistency rate of $\sim$90$\%$. 
Using these visual classifications in combination, we estimated the precision of the classifier ($p'$) as a function of varying score threshold by:

\begin{equation}
p'(x)=\frac{N_1(s>x)}{N_1(s>x)+N_0(s>x)},
\end{equation}

\noindent where $N_1$ and $N_0$ are the number of true and false positives according to human experts, $s$ is the classifier score, and $x\in\mathbb{R}[0,1]$. 
Figure~\ref{fig:precision} shows the resulting $p'(x)$ curve from our visual validation process, together with the kernel density estimates of true positives, false positives, and all data for our {\em broad selection}.
We set our decision boundary to $s=0.6$, which corresponds to a human-estimated precision of $p'=0.9$. 
The objects that were visually classified as RRL but with $s<0.6$ have typically fewer data points, noisy light curves, and/or strong systematics due to, e.g., source crowding.
Finally, we note that standard measures of classifier performance, such as the $F_1$ measure or the receiver operating characteristic (ROC) curve, cannot be estimated by human visual inspection because they require the measurement of false negatives, which cannot be obtained in the absence of previously labeled data (i.e., known RRab stars in the studied area).

Using a score threshold of 0.6 resulted in a {\em narrow selection} of 2147 RRab candidates. 
Subsequently, we flagged all objects that were classified as non-RRab by at least one human expert. 
By this last step, we further increased the precision in our sample to probably a few percent, at the expense of lowering the recall by an unknown (but presumably tiny) amount. 
The unflagged objects with $s>0.6$ form our final, {\em fiducial sample} of 1892 objects, which will serve as the basis of our analyses in Sections~\ref{sec:charact}--\ref{sec:map}.

At a first glance, the exclusion of all flagged objects from our final sample might seem a quite drastic and perhaps too conservative step because the gain of precision might cost a significant decrease of sample completeness. 
We argue that this is not the case because an inspection of the flagged objects reveals that the human experts had quite objective reasons for flagging in general, namely suspected blending, large systematics in the photometry, low $S/N$, and signatures that the candidate is a contact eclipsing binary. 
The confusion with binary light curves is a particularly challenging aspect of RRab classification based on near-IR data. 
This is because some RRab light curves can be symmetric, and resemble the phase diagrams of contact eclipsing binaries phase-folded with half of their true periods. Figure~1 of \cite{2016A&A...595A..82E} displays an excellent illustration of this issue. 
An increased local scatter around minimum brightness is a good signature of a binary being a false positive (due to the different depths of the primary and secondary minima), and our classifier learned to distinguish binaries from RRab stars based on such signatures via the {\tt p2p\_scatter\_2praw} feature (see \citealt{2016A&A...595A..82E} for its definition). 
However, due to the sparser photometric sampling of the VVV disk fields, this important signature can be quite subtle in some cases, in which the judgment of a human expert's eye, although not a deterministic inference, can sometimes be more accurate than a machine-learned classifier.

\begin{figure}[]
\includegraphics[width=0.45\textwidth]{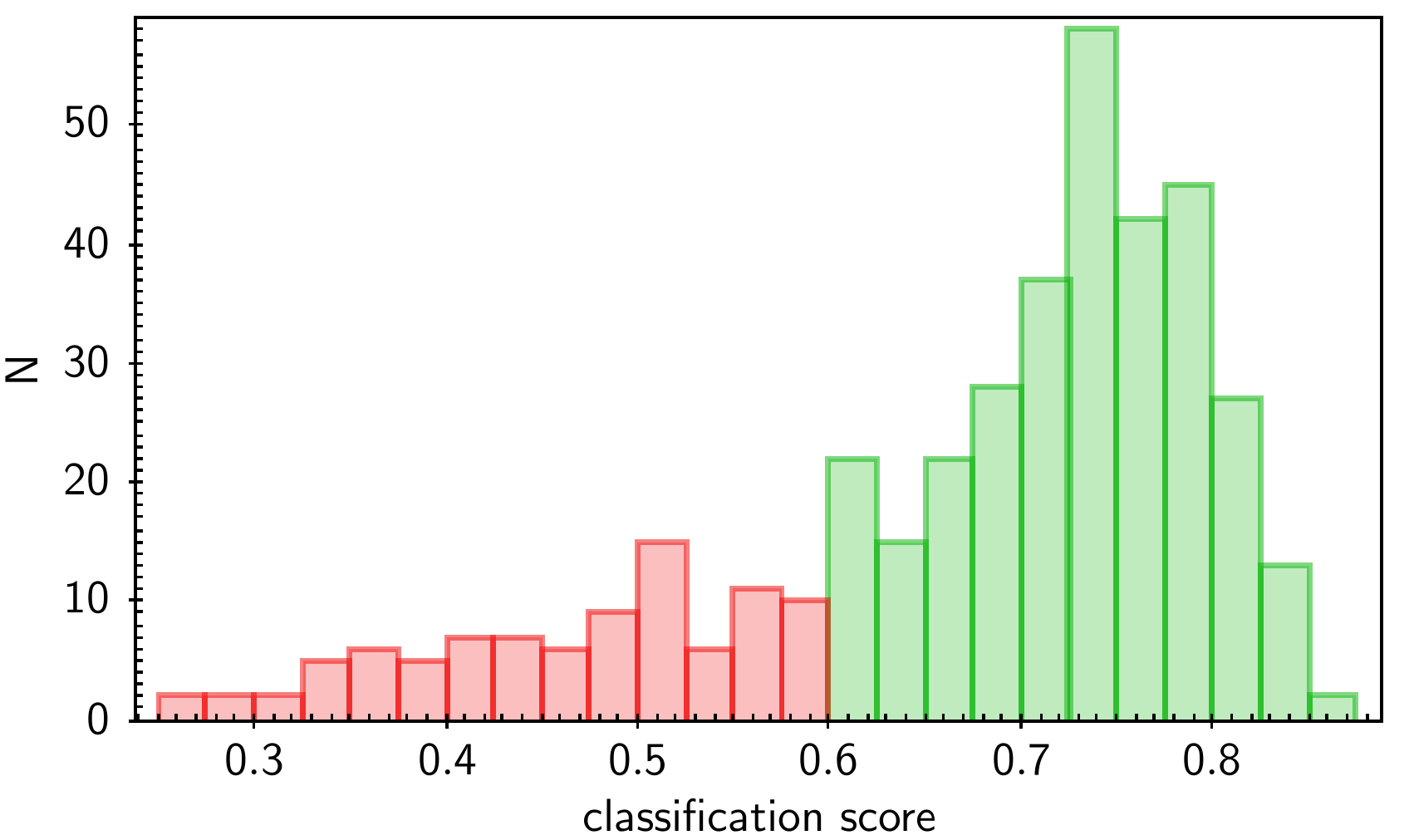}
\caption{Histogram of the classifier scores of the 404 RRab candidates in the study by \cite{2017AJ....153..179M}. Data above and below our decision boundary are colored green and red, respectively.\label{fig:ajpap_rr}}
\end{figure}

We compared our results to the catalog of 404 RRab candidates in the earlier study of \cite{2017AJ....153..179M}, which covered 25$\%$ of the total disk area of the VVV survey. 
All objects were included in all steps of our present analysis, and their classification scores are shown in Figure~\ref{fig:ajpap_rr}. 
Only $\sim$77$\%$ of the previous candidates, i.e., 312 objects, have scores above our current threshold, and 295 of them were not flagged in our analysis (i.e., included in our fiducial sample). 
We note that some of the candidates in \cite{2017AJ....153..179M} got extremely low scores by the machine learned classifier. 
On the other hand, our narrow selection contains 673 (593 unflagged) objects in the region covered by \cite{2017AJ....153..179M}, 361 (298 unflagged) of which were not included in the \cite{2017AJ....153..179M} sample.

Although both studies involved visual inspection of light curves for object classification, there is a substantial difference between their methodologies, which resulted in such a  different performance. In the case of our present study, the human component was limited to objects with high classifier scores, it was employed with high redundancy in order to minimize personal bias, and its purpose was to only fine-tune the decision boundary of a deterministic method, as well as to slightly increase the purity of the final sample. On the other hand, the object classification by \cite{2017AJ....153..179M} was completely based on the visual inspection of $\sim$10$^4$ light curves by a single human, and included very limited redundancy only in the inspection of the best-looking $\sim$700 candidates, leading to poor performance. We conclude that the visual classification of near-IR data is very risky, especially when performed on extremely large data sets without redundancy, because fatigue will lead to a high error rate and a time-varying decision boundary in the human brain, thus resulting in both incomplete and contaminated samples, which can easily lead to false scientific conclusions. We stress that a purely visual classification of light curves in large near-IR time-domain surveys is highly discouraged.

\section{Near-IR photometric characterization of the RR Lyrae stars} \label{sec:charact} 

A great advantage of using RRL stars as population tracers is that their reddenings, distances, and metallicities can be precisely estimated from parameters derived from photometric measurements. 
The simple scheme of RRL parameter estimation from 2 photometric bands is as follows.

\begin{enumerate}

\item The [Fe/H] heavy element abundance of the star is estimated from its light curve properties. Exactly which light curve features are employed in this prediction depends on the photometric band.

\item The estimated [Fe/H] values are converted into absolute heavy element content $Z$. For this conversion, it is necessary to make an assumption for the star's helium content and adopt a heavy element mixture.

\item The period-luminosity-metallicity relations are used to predict the $M_i$ absolute magnitude of an RRL star in the IR photometric band $i$ :
\begin{equation}
M_i=f(\log{P},\log{Z}),
\end{equation}
where $P$ is the pulsation period, $Z$ is the total heavy element content of the object, and $f$ symbolizes the functional form of the relation.

\item The color excess (i.e., reddening) $E_{ij}$ of the star in the $i$ and $j$ passbands is then calculated as:
\begin{equation}
E_{ij}=\langle m_i\rangle-\langle m_j\rangle-(M_i-M_j),
\end{equation}
where $\langle m_i\rangle$ is the observed {\em mean magnitude} of the star in the $i$ band.

\item The absolute extinction $A_i$ is then estimated from $E_{ij}$, based on the total-to-selective extinction ratio implied by an assumed reddening law.

\item Finally, the distance $d$ (in pc) to the object is calculated by:
\begin{equation}
\log{d} = 1 + 0.2 \cdot (m_i - A_i - M_i).
\end{equation}
\end{enumerate}

In the following, we discuss the details of the various steps of this procedure.

\subsection{Metallicity estimation} \label{sec:feh}

The empirical relationships between the metallicity and the optical light curve shape of RRL stars \citep[e.g.,][]{1995A&A...293L..57K,1996A&A...312..111J,2005AcA....55...59S} enable us to employ them as metallicity tracers. 
However, an empirical calibration linking spectroscopic metallicities to the near-IR light curve parameters of RRLs has been lacking, despite its key importance in the era of large time-domain photometric surveys like the VVV, since a large fraction of the newly discovered distant RRLs are beyond the faint magnitude limit of optical surveys. 
This situation arises from the fact that as of now, RRL stars with both spectroscopic abundance measurements and well-sampled near-IR light curves have not been available in sufficient numbers. 
An important step forward in this context was taken by the discovery of a nonlinear relationship between the $K_s$-band light curves of RRab stars and the photometric metallicities derived from their $I$-band counterparts using the relation of \citet[Eq. 3]{2005AcA....55...59S}. 
In \cite{hajdu}, we trained a predictive model of the metallicity using a large number of bulge RRab stars discovered by the OGLE-IV survey in the optical $I$ band, which also have accurate $K_s$-band light curves acquired by the VVV survey. 
A detailed description of the method is provided in \cite{hajdu}, and here we only briefly summarize the main properties of this estimator.

Firstly, the $K_s$ light curves in the training data set of this method were represented by the linear combination of the Fourier parameters of RRab principal component (PC) light curves, instead of using the traditional direct Fourier fitting. 
The PCs were computed from accurate Fourier models of a large representative set of high-quality RRab light curves from the VVV survey, and result from the singular value decomposition of their phase-folded and phase-aligned light curves, with their standard-scaled magnitudes in 100 equidistant phase points used as input features. 
\cite{hajdu} found that any RRab light curve can be accurately modeled with a linear combination of the Fourier parameters of the first 4 PCs. 
The main advantage of this method when compared to standard Fourier fitting is that it has many fewer parameters, namely the period and the coefficients of the PCs ($U_i$, $i=\{1\dots4\}$), which makes it more robust in case of relatively few data points or when the light curve is affected by phase gaps. 
Subsequently, the PC-based light curve model was decomposed into a truncated Fourier-series, in order to also provide a standard representation of the data.

The prediction formula of [Fe/H] values was determined from the period and the amplitudes and phases of the first three Fourier terms via regularized nonlinear regression performed by a small neural network (see \citealt{hajdu} for details). The mean prediction accuracy with respect to the \cite{2005AcA....55...59S} calibration is 0.2~dex in the [Fe/H]$\in\mathbb{R}[-1.8,0]$~dex range. This was computed via standard cross-validation procedures and was further verified on various independent test datasets, which are discussed in \cite{hajdu}.

Concerning the accuracy, it is important to make a note of the deficiency of low-metallicity objects in the spectroscopic calibrating samples of both the $V$- and $I$-band [Fe/H] formulae (i.e., the calibrators behind our training sample), which leads to a bias in the estimated metallicities. 
For example, the [Fe/H] values predicted by the two-parameter $V$-band formula are known to be biased for stars with ${\rm [Fe/H]} < -2$ \citep{1996A&A...312..111J}. 
\cite{hajdu} have shown that the 3-parameter $I$-band formula of \cite{2005AcA....55...59S} provides significantly more unbiased [Fe/H] estimates for low-metallicity, Oosterhoff II RRab stars (\citealt{1939Obs....62..104O}, see also \citealt{2015pust.book.....C}, for a recent review and references), and made further small corrections in the training sample. 
Since our de-biased training sample encompasses a very wide range of metallicities, our [Fe/H] estimates determined from $K_s$-band photometry are free from any strong regression bias, except for the extremely metal-poor tail of the distribution at [Fe/H]$\lesssim-1.8$. At the same time, we note that our regression residual may carry some weak nonlinearities at the $\sim0.05$~dex level \citep[see][]{hajdu}.

The total accuracy of the [Fe/H] values with respect to the spectroscopic measurements based on which the \cite{2005AcA....55...59S} formula was calibrated, is $\sim$0.25~dex, considering the prediction error of the latter. 
We note that the predicted [Fe/H] values are on the \cite{1995AcA....45..653J} metallicity scale, which is based on high-dispersion spectroscopic (HDS) measurements. 
A more recent HDS-based metallicity scale was established by \citet[hereafter C09]{2009A&A...508..695C}, using a large number of red giant stars in globular clusters. 
Since their measurements do not include RRLs, it is not possible to directly calibrate photometric metallicity estimates to the C09 scale.
However, a linear transformation of our predicted [Fe/H] values to the C09 scale is possible \citep{hajdu}. 
The offset between the two scales is [Fe/H]-dependent, and ranges approximately between 0.02 and 0.1 dex.

\begin{figure}[]
\includegraphics[width=0.48\textwidth]{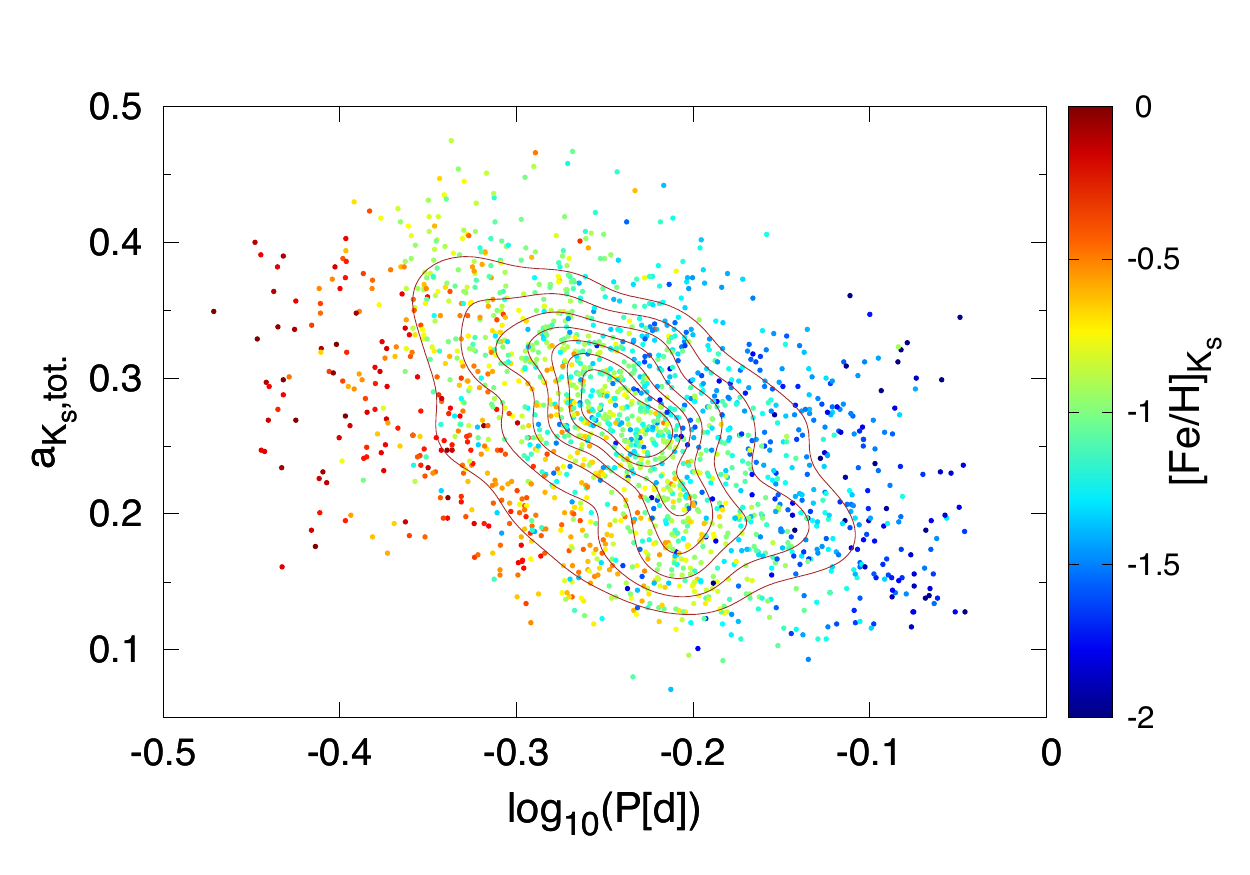}
\caption{Near-infrared Bailey ($K_s$ total amplitude {\em vs} period) diagram of the RRLs in our study. The contours represent a Gaussian kernel density estimate of the points with an optimal kernel size, determined by leave-one-out cross-validation, in order to highlight the main locus of stars in this diagram. The color scale shows the [Fe/H] values of the individual stars estimated from the $K_s$-band light curves.\label{fig:logp-atot-feh}}
\end{figure}

We computed photometric metallicities for our fiducial sample of disk RRab stars by first performing a PC regression of the post-processed $K_s$-band light curves, and then applying the \cite{hajdu} method on the fitted parameters. Figure~\ref{fig:logp-atot-feh} shows our fiducial sample of RRab stars on the Bailey diagram with their metallicities color-coded, and their number density estimate highlighted with contours. A relatively large number of metal-rich stars, as well as metal-poor Oosterhoff II objects, are located at the left and right sides of the main locus, respectively. The resulting RRL MDF will be discussed in detail in Sect.~\ref{sec:map}.

The absolute heavy element content $Z$ was calculated from the [Fe/H] values by assuming a helium content of $Y=0.245$ and an alpha-element enhancement of $[\alpha/{\rm Fe}]=0.3$, standard solar chemical composition \citep{1998SSRv...85..161G}, and chemical element distributions at $[\alpha/{\rm Fe}]\neq0 $ provided by \cite{2005ApJ...623..585F}. We first computed a grid of [Fe/H] {\em vs} $\log{Z}$ values, and obtained the following formula by linear regression:
\begin{equation}
\log{Z}={\rm [Fe/H]} - 1.538,
\end{equation}
which was used for converting [Fe/H] to $\log{Z}$.

\subsection{Extinctions and distances}

\begin{figure*}[]
\centering
\includegraphics[width=0.8\textwidth]{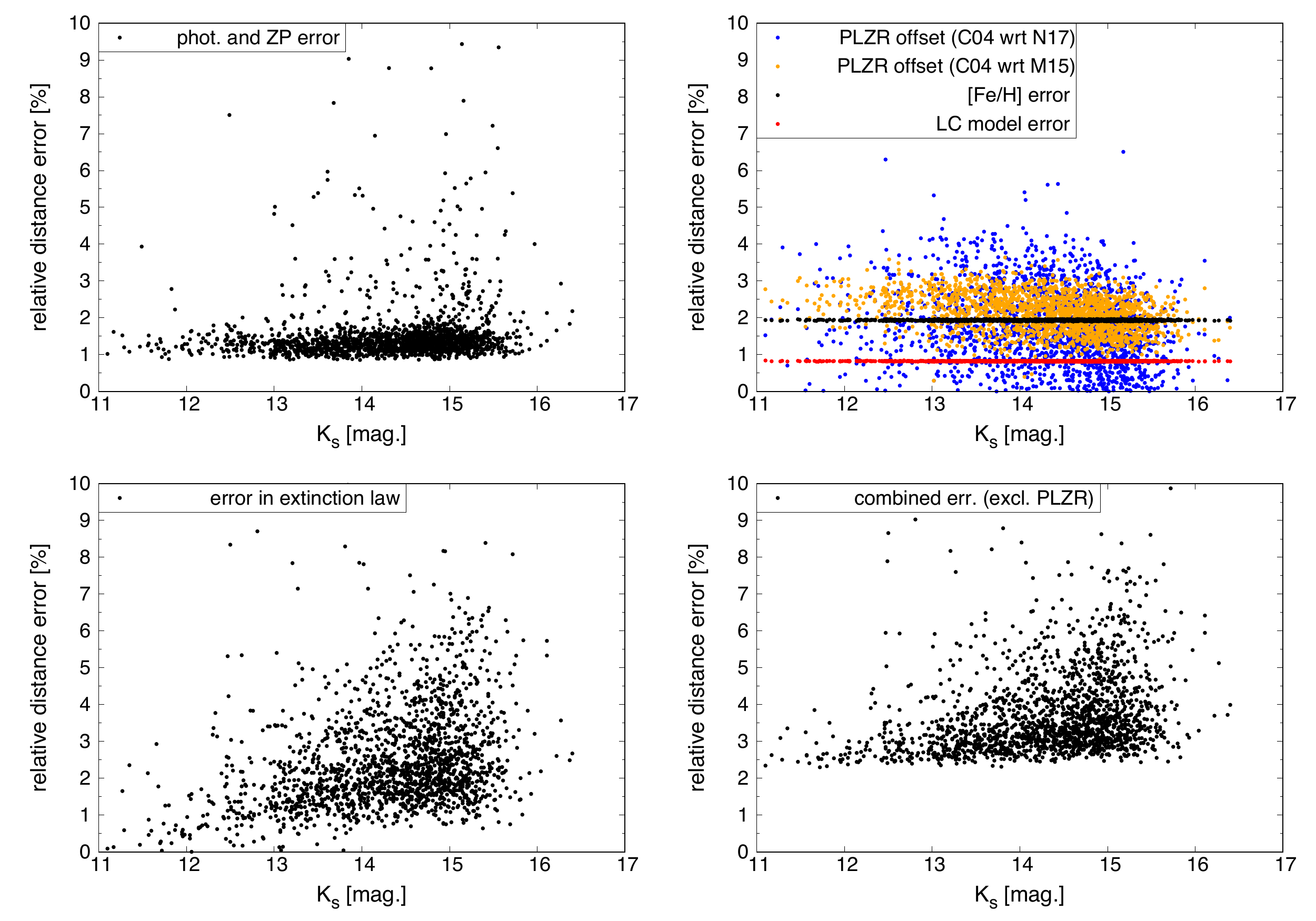}
\caption{Relative errors in the distances to the RRLs (based on the $J$ and $K_s$ measurements) originating from different sources, as a function of the apparent $K_s$ magnitude. {\em Top left:} errors propagated from the photometric and magnitude zero-point uncertainties. {\em Bottom left:} errors originating from the nominal uncertainty of the \cite{2016A&A...593A.124M} mean selective-to-total extinction ratio. {\em Top right:} errors arising from the error in the [Fe/H] estimates {\em (black)}, from the light curve models {\em (red)}, and the relative offset in distances when using the PLZRs of \citet[][N17]{2017A&A...604A.120N} and \citet[][M15]{2015ApJ...808...50M} instead of those from \citet[][C04]{2004ApJS..154..633C} ({\em blue} and {\em orange}, respectively). {\em Bottom right:} combined errors, excluding the uncertainty in the PLZRs. \label{fig:errors}}
\end{figure*}

The absolute magnitudes of the RRLs were estimated using the theoretical PLZRs of \cite{2004ApJS..154..633C}. Since the formulae in their original study were computed for different photometric systems, we first had to convert those into the VISTA system. This was performed by transforming the $JHK$ magnitudes for each and every individual synthetic star in the \cite{2004ApJS..154..633C} simulations into the VISTA photometric system, and then recomputing all linear regressions. The magnitudes were converted using eqs. (A1)-(A3) of \cite{2001AJ....121.2851C}, and the formulae provided by the Cambridge Astronomy Survey Unit\footnote{\texttt{http://casu.ast.cam.ac.uk/surveys-projects/vista/\\technical/photometric-properties/}}. The resulting period--luminosity--metallicity relations in the VISTA system are as follows:

\begin{eqnarray}
M_{K_s} & = & -0.6365 - 2.347 \cdot \log{P} + 0.1747 \cdot \log{Z}\,, \\
M_H  & = & -0.5539 - 2.302 \cdot \log{P} + 0.1781 \cdot \log{Z}\,, \\
M_J  & = & -0.2361 - 1.830 \cdot \log{P} + 0.1886 \cdot \log{Z}\,.
\end{eqnarray}

The reddening was estimated by the comparison of the observed intensity-averaged color index to the intrinsic theoretical equilibrium color index, computed from the PLZRs, independently for each star.

RRLs undergo large temperature changes during their pulsation cycles, hence their color indices change significantly as a function of pulsation phase. 
Since the VVV survey provides only a very limited number of $J$ and $H$ observations, a direct unbiased estimation of the mean color indices is not possible. We mitigated this problem by estimating the mean $J$ magnitude by means of a machine-learned predictive model discussed by \cite{hajdu}. This model is capable of predicting the relative variations of the $J$ magnitude as a function of the pulsation phase (i.e., the shapes of the light curves) from the model parameters of the $K_s$ light curve with a very high accuracy of $\sim$0.02~mag. The mean $H$ magnitudes were computed under the assumption that an RRL star's light curve shape is identical in the $H$ and $K_s$ bands \citep[for details, see][]{hajdu}. The predicted light curves are then fitted to the actual $J$- and $H$-band measurements (since their pulsation phases are known), and the intercepts are used as estimates of $\langle J \rangle$ and $\langle H \rangle$.

Distances were computed from the distance moduli measured in the $K_s$ filter because among the available photometric wavebands, the mean magnitude is most accurate, the metallicity dependence and the intrinsic scatter of the PLZR are the smallest, and the extinction is the lowest in this band. 
The absolute $K_s$-band extinction of each star was inferred from the $E(J-K_s)$ or (in the absence of a $J$-band detection) the $E(H-K_s)$ color excess, using the mean selective-to-total extinction ratios of \cite{2016A&A...593A.124M}, namely $A(K_s)/E(J-K_s)=0.49$ and $A(K_s)/E(H-K_s)=1.49$.
The advantage of employing the \cite{2016A&A...593A.124M} relations is that they were measured using RRLs and Cepheids, therefore they are free from biases carried by other relations that rely on proxies with significantly different spectral energy distributions, e.g., red clump (RC) stars. 
Moreover, these relations were obtained using objects lying at typically low Galactic latitudes, thus our estimates are less affected by a possibly significant spatial variation in the extinction curve between high- and low-latitude sight-lines, as found for optical--near-IR extinction ratios \citep[e.g.,][]{2016MNRAS.456.2692N}. 
At the same time, we note that the existence of such variations is currently debated \citep[see, e.g.,][]{2016ApJ...821...78S}, and our analysis would be only marginally affected by them (see Sect.~\ref{sec:map}).

\begin{figure*}[]
\centering
\includegraphics[width=\textwidth]{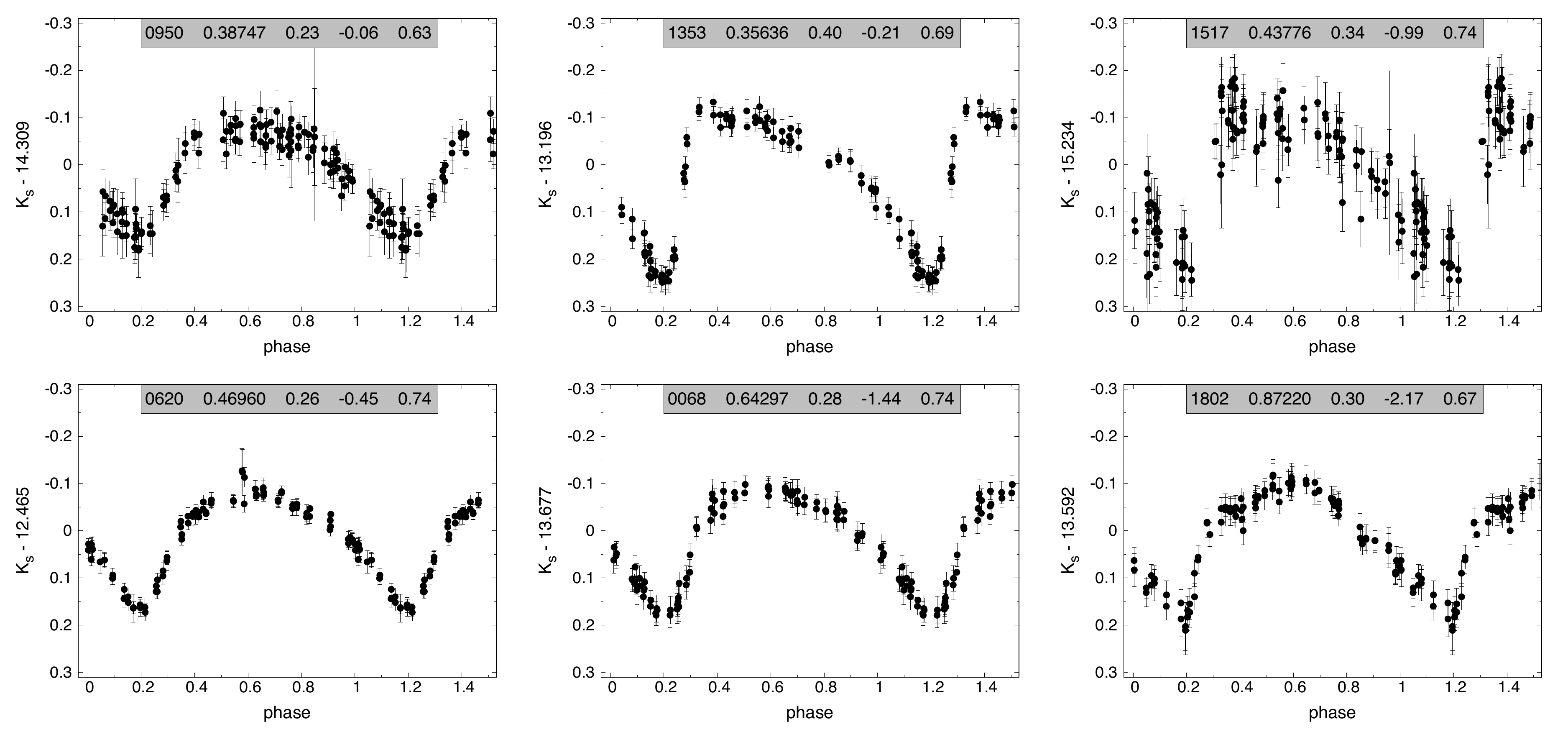}
\caption{Phase-folded light curves in the $K_s$ band of RRLs from our fiducial sample. The phase point 0.2 was defined as the phase of the minimum brightness for clarity. The numbers in the figure headers from left to right are as follows: object identifier, period (in days), total amplitude, photometric [Fe/H] estimate, and classification score.\label{fig:lcplot}}
\end{figure*}

\begin{figure*}[]
\centering
\includegraphics[width=\textwidth]{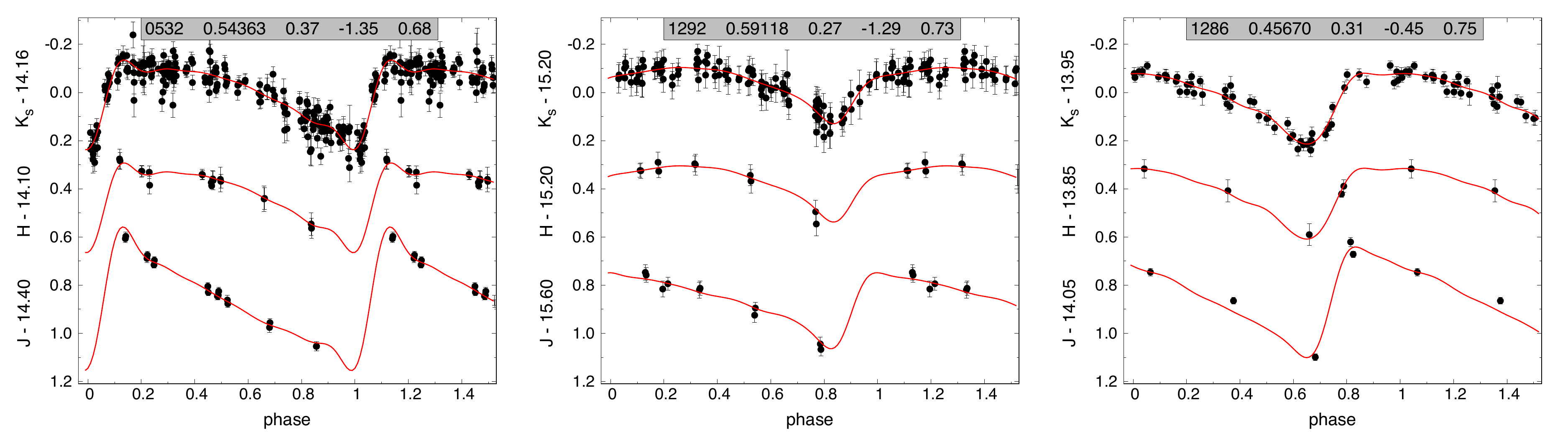}
\caption{Phase-folded light curves in the $J, H, K_s$ bands of RRLs from our fiducial sample, from fields with the highest number of $J$- and $H$-band observations. The red lines show the PC fits to the $K_s$ light curves, and the predicted $J$ and $H$ light curves with their intercepts fitted to the measurements. The numbers in the figure headers have the same meaning as in Fig.~\ref{fig:lcplot}.\label{fig:jhkplot}}
\end{figure*}

\begin{table*}[h!]
\caption{Parameters of the RRLs in Our {\em Fiducial Sample}. \label{tab:photpars}}
\begin{splittabular}{ccccccccccBccccccccc}
\hline 
\hline 
 ID &  R.A. [h:m:s] &       Decl. [d:m:s]   &   Score\tablenotemark{a} &  Period~[d]  &  ap.\tablenotemark{b} &   $\langle K_s \rangle$\tablenotemark{c}  &  $\sigma$\tablenotemark{d}  &  $S/N$  &
   $a_{K_s,{\rm tot.}}$\tablenotemark{e} &  $\langle J \rangle$\tablenotemark{c} &  $\langle H \rangle$\tablenotemark{c} &  $U_1$\tablenotemark{f}  &  $U_2$\tablenotemark{f} &          $U_3$\tablenotemark{f} &  $U_4$\tablenotemark{f} &
     [Fe/H]$_{K_s}$ &   ${\rm E}(J-K_s)$  &    $d$ [kpc]  \\
\hline
  1   &   11:38:55.83  &  -63:22:37.8  &  0.778  &  0.456029  & 3  &  13.966  &  0.019  & 128.4  &  0.254  &  15.009  & 14.165  &  0.774  &   0.134  &  0.002    &  -0.005  &   -0.28  &  0.844  &  5.51    \\    
  2   &   11:40:13.71  &  -62:34:28.7  &  0.771  &  0.525924  & 2  &  15.007  &  0.049  &  68.3   &  0.342  &  16.971  & 14.815  &  1.007  &  -0.158  & 0.009     &  -0.048  &   -0.95  &  1.742  &  8.20    \\  
  3   &   11:41:08.07  &  -62:57:22.4  &  0.734  &  0.627923  & 3  &  14.122  &  0.019  & 108.1  &  0.213  &  15.257  & 14.398  &  0.638  &   0.237  &  -0.035   &  -0.044  &   -0.94  &  0.874  &  7.21    \\       
  4   &   11:41:19.57  &  -62:46:45.3  &  0.663  &  0.898821  & 4  &  13.517  &  0.017  & 103.4  &  0.187  &  14.401  & 12.924  &  0.556  &   0.306  & -0.033    &  -0.047  &   -1.64  &  0.552  &  7.34    \\     
  5   &   11:42:24.58  &  -62:00:41.2  &  0.670  &  0.507736  & 3  &  14.607  &  0.026  &  93.7   &  0.212  &  15.627  & 14.941  &  0.583  &   0.235  &  -0.040   &  -0.050  &   -0.41  &  0.799  &  7.95    \\      
\hline
\end{splittabular}
\tablecomments{The full table is available in machine-readable format in the online edition of the paper.}
\tablenotetext{a}{Classification score (see Sect.~\ref{subsec:classif}).}
\tablenotetext{b}{Optimal aperture (see Sect.~\ref{subsec:per}).}
\tablenotetext{c}{Magnitudes of the intensity means.}
\tablenotetext{d}{Standard deviation of the residual.}
\tablenotetext{e}{Total amplitude in the $K_s$ band.}
\tablenotetext{f}{Amplitudes of the first four principal components (see Sect.~\ref{sec:feh} and \citealt{hajdu}).}
\end{table*}

\begin{deluxetable}{cccccc}
\tablecaption{$JHK_s$ photometric time-series of the candidate RRLs with ${\rm Score}>0.6$. \label{tab:phot}}
\tablehead{
\colhead{ID} & Filter & \colhead{Julian Date\tablenotemark{a}} & mag. & phot. err. & ZP err.
}
\startdata
1 & $K_s$ & 55938.846162 & 14.056 & 0.031 & 0.024 \\
1 & $K_s$ & 55938.847161 & 14.067 & 0.021 & 0.010 \\
1 & $K_s$ & 55947.765248 & 13.923 & 0.028 & 0.028 \\
1 & $K_s$ & 55947.766437 & 13.931 & 0.026 & 0.013 \\
\multicolumn{4}{l}{\dots} \\
\enddata
\tablenotetext{a}{${\rm HJD}-2455000.0$ are given.}
\tablecomments{The full table is available in machine-readable format in the online edition of the paper.}
\end{deluxetable}

Errors in the distances have several sources, and in the following, we provide a brief characterization of each of them. 
Photometric errors and ZP errors (determined for each exposure) cause uncertainties in the mean magnitudes. 
The photometric errors have been discussed in detail by \cite{2012A&A...537A.107S}, and range typically from 0.01 for bright stars up to 0.1 magnitudes for our faintest RRLs, near the detection limit; while ZP errors usually fall between 0.01 and 0.03 magnitudes. 
Both have significant contributions in the estimated $\langle J \rangle$ and $\langle H \rangle$, but ZP errors dominate in $\langle K_s \rangle$ since there are up to 50 epochs in that band. 
Since the mean magnitudes are computed via model fits, the inaccuracy of the latter brings in additional uncertainties. 
$\langle K_s \rangle$ is the intercept of the fitted Fourier series, and its error is smaller than 0.01 mag. 
$\langle J \rangle$ and $\langle H \rangle$ are computed from predictive models, which have a mean statistical error of 0.02 magnitudes in both bands \citep[see][]{hajdu}.

Further errors propagate in through the PLZRs: errors in the estimated [Fe/H] values, possible systematic errors in the $\log{Z}$ conversion, and the uncertainties in the PLZ relations themselves. 
The latter can be estimated by comparing distance estimates using different (empirical or theoretical) period-luminosity relations. 
In the case of empirical relations, this is complicated by several factors: their uncertainties are dominated by either small sample sizes or small [Fe/H] ranges and different metallicity scales; they are measured in various different photometric systems; and empirical near-IR PLZRs are scarcely derived for bands other than $K_s$, which would result in a blending of distance errors arising from the PLZRs with those arising from the extinctions when lacking a star-by-star estimate of the intrinsic color. 
A detailed systematic comparison of PLZRs is beyond the scope of this paper, thus we point the reader to the recent works of \cite{2015ApJ...807..127M} and \cite{2016ApJ...832..210B}. 
In order to give a reasonably good estimate of the systematic error arising from the uncertainty in the PLZRs, we compared the distances to the RRL stars to those obtained by using the empirical $J$ and $K_s$ PLZRs of \cite{2017A&A...604A.120N}, which are based on the VISTA time-series photometry of RRab stars in $\omega$\,Centauri, and thus span a relatively large range in metallicity. Moreover, we also compare our distances to the theoretical PLZRs to the ones obtained by \cite{2015ApJ...808...50M}, which are based on different pulsation models and evolutionary constraints compared to those of \cite{2004ApJS..154..633C}.

Figure~\ref{fig:errors} exhibits the contribution of relative distance offsets and errors propagating from various sources of uncertainties, computed independently for each RRL by Monte Carlo simulations. 
The errors caused by the inaccuracy in the photometry, [Fe/H] estimation, and light curve models do not exceed the 1-2\,\% level for the vast majority of the objects. 
The systematic distance offset between the solutions obtained by the \cite{2004ApJS..154..633C} PLZRs with respect to the \cite{2017A&A...604A.120N} and \cite{2015ApJ...808...50M} PLZRs is also on the 1--2\,\% level (depending mainly on the period). We note that the results presented later in this paper are unaffected by our use of these other PLZRs. The impact of the extinction ratio's uncertainty on the distances is larger roughly by a factor of 2, and is identified as the most important source of error, although its effect should be largely systematic, since a large variation in the near-IR part of the extinction curve over small angular scales is unlikely, according to the latest studies \citep[see, e.g.][]{2016ApJ...821...78S,2016A&A...593A.124M}. 
The combined relative distance errors (excluding the uncertainty in the PLZR) are typically 3--4\,\%, with $\sim$90\,\% of the stars falling below the 5\,\% level.

Table~\ref{tab:photpars} presents the coordinates, periods, magnitudes, light curve parameters, various descriptive statistics, classification scores, [Fe/H] estimates and distances for all 1892 objects in our fiducial sample. Figure~\ref{fig:lcplot} shows a small representative sample of $K_s$-band light curves from our fiducial sample with various periods and metallicities. Figure~\ref{fig:jhkplot} shows the $JHK_s$ light curves of three of our objects with a relatively large number of $J$- and $H$-band observations compared to the rest of the sample, together with the PC fit in the $K_s$ band and the predicted light curves in the $J$ and $H$ bands (the latter being identical to the $K_s$-band fit), fitted to the data points \citep[for details, see][]{hajdu}. The complete data set that serves as the basis of this study is available in the online edition of the journal. Table~\ref{tab:phot} contains the full time-series photometry for all objects in our fiducial selection in the VISTA $JHK_s$ photometric system.

\section{Spatial and metallicity distribution} \label{sec:map}

\subsection{Three-dimensional distribution}

\begin{figure*}[h]
\centering
\includegraphics[width=0.8\textwidth]{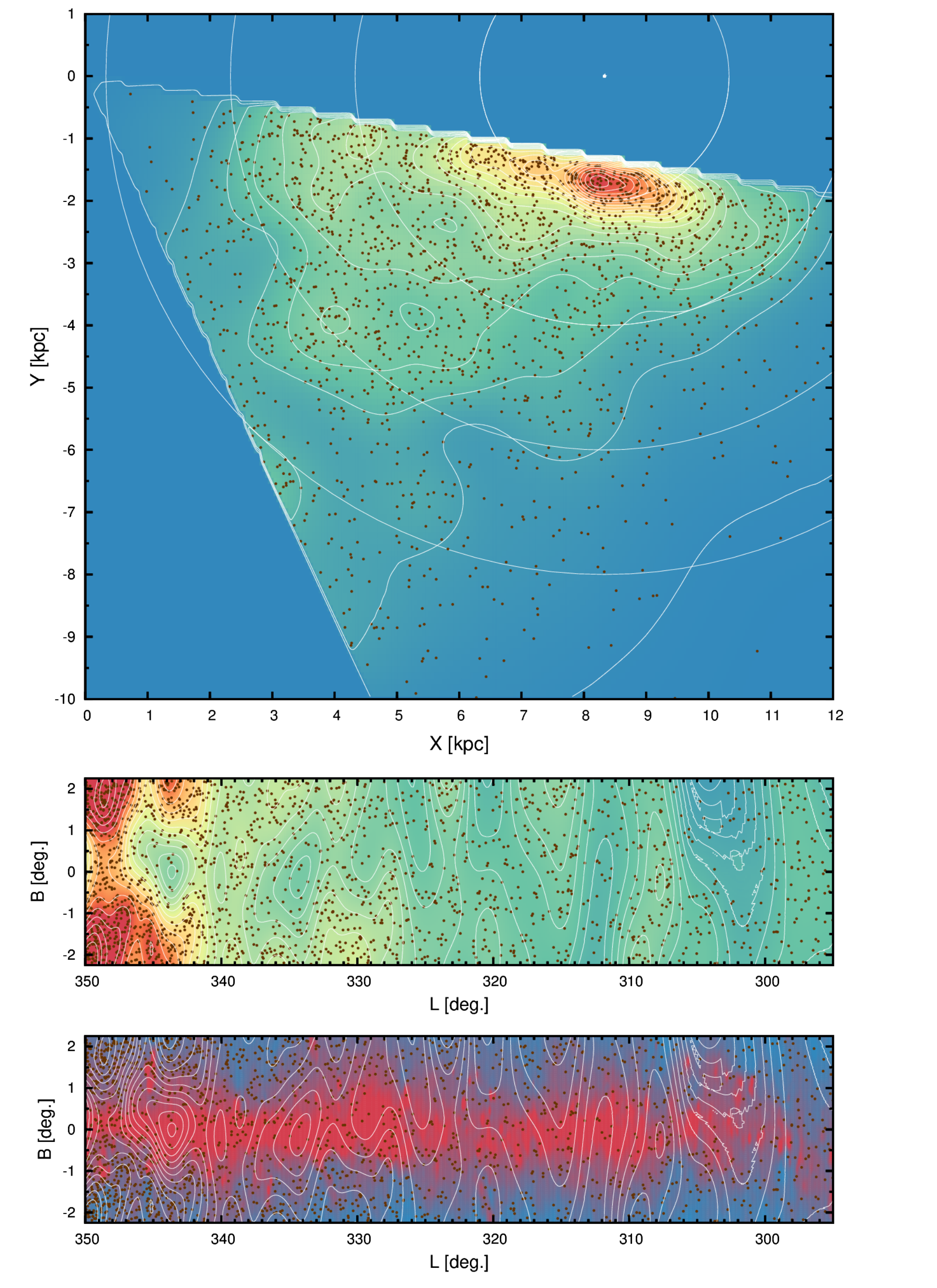}
\caption{{\em Top:} spatial distribution of the RRab stars projected onto the Galactic plane. The brown dots show the position of individual objects, and the color scale and contour lines represent a kernel density estimate with a kernel size of 400 pc, with a linear density increment between contours. The concentric circles mark Galactocentric cylindrical distances in 2\,kpc increments, whereas the white point marks the Galactic Center. {\em Middle:} same as above but for the positions of the objects on the sky (in Galactic coordinates), using a kernel size of $0.75^\circ$, and with logarithmic density increment between contours. {\em Bottom:} same as above but the color scale represents the reddening map of \cite{1998ApJ...500..525S} in the $1<A(V)<15$ range. \label{fig:spacedist}}
\end{figure*}

The spatial distribution of the RRLs projected on the Galactic plane and on the sky are shown in the top and middle panels of Fig.~\ref{fig:spacedist}, respectively. 
It is immediately apparent that the distributions are dominated by the selection function of the VVV survey, in particular the variations in the detection efficiency due to differential extinction. 
The global maximum in the density corresponds to the outskirts of the bulge, and its Galactic $X$ coordinate matches the dynamical estimate of the distance to the Galactic Center \citep[]{2009ApJ...692.1075G}, which confirms that the outskirts of the bulge RRL distribution are not barred \citep[e.g.,][]{2013ApJ...776L..19D,2017AJ....153..179M}. 
At $l\lesssim345^\circ$, the source density drops quickly and its variations become dominated by extinction variations.  The bottom panel of Fig.~\ref{fig:spacedist} compares the RRL density with the distribution of optical extinction on the sky based on the reddening map of \cite{1998ApJ...500..525S}, showing that the small apparent overdensities qualitatively follow areas of lower extinction. The projected Galactic ($X,Y$) distribution is also free from any immediately evident structure other than the bulge and those caused by extinction variations. A more detailed analysis of subtle features in the spatial density would require the quantitative knowledge of the three-dimensional RRL detection efficiency function of the VVV survey, and is beyond the scope of this study.

\begin{figure*}[]
\centering
\includegraphics[width=\textwidth]{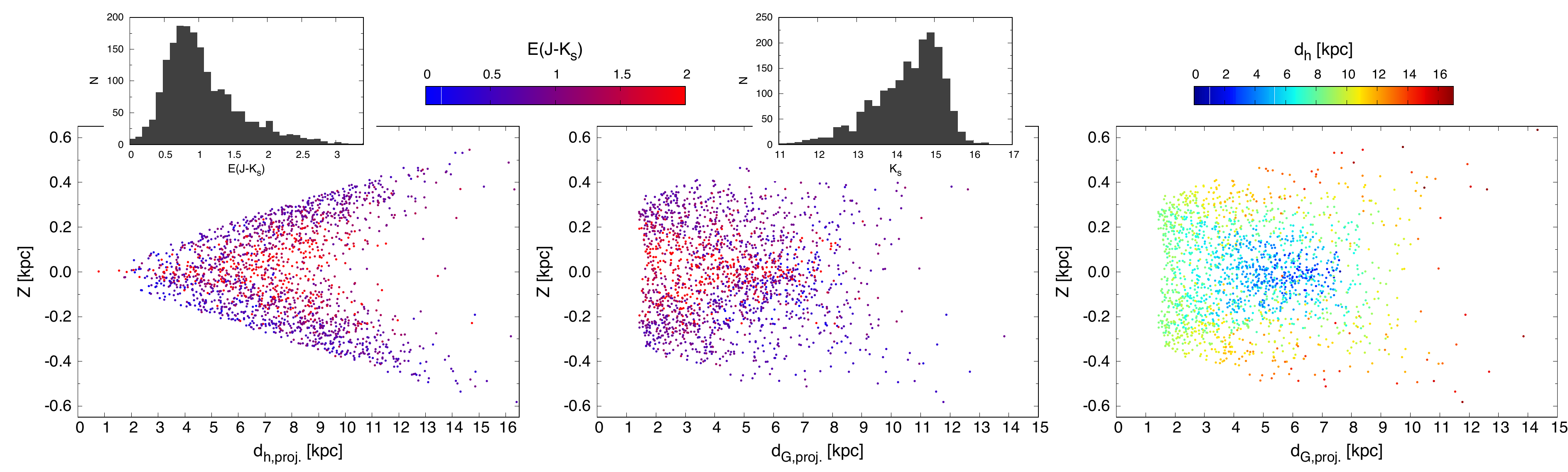}
\caption{{\em Bottom left:} marginalized heliocentric distance distribution of the RRLs. The color scale represents the $E(J-K_s)$ color excesses of the stars. {\em Bottom middle:} Same as above, but marginalized around the Galactic Center. {\em Bottom right:} same as the bottom middle panel, but the color scale represents the heliocentric distances of the objects. {\em Top left:} histogram of the reddening values of the RRLs. {\em Top middle:} histogram of the apparent mean $K_s$ magnitudes of the RRLs. \label{fig:dists}}
\end{figure*}

We computed the Galactocentric distance distribution of the RRLs assuming that the distance to the Galactic Center is $R_0=8.3$\,kpc \citep{2009ApJ...692.1075G,2016ApJS..227....5D}. 
Figure~\ref{fig:dists} compares the heliocentric and Galactocentric marginalized spatial distributions. 
The former is heavily biased by selection effects mainly due to extinction: RRLs are detected up to distances of $\sim$16\,kpc toward the high-latitude edges of the VVV disk footprint, but objects lying in very close proximity of the Galactic plane remain undetected beyond $\sim$8--9\,kpc. 
Another, smaller selection effect is a decreased detection rate of objects at short heliocentric distances: due to the saturation limit of the VVV survey at around $K_s \lesssim 11$, the closest RRLs can remain undetected in the absence of sufficient extinction. 
The lack of a marked truncation in the distribution at short distances is due to the observational cone: objects closer to the Sun are closer to the Galactic plane as well, therefore their extinctions tend to be systematically higher. 

\begin{figure}[]
\includegraphics[width=0.48\textwidth]{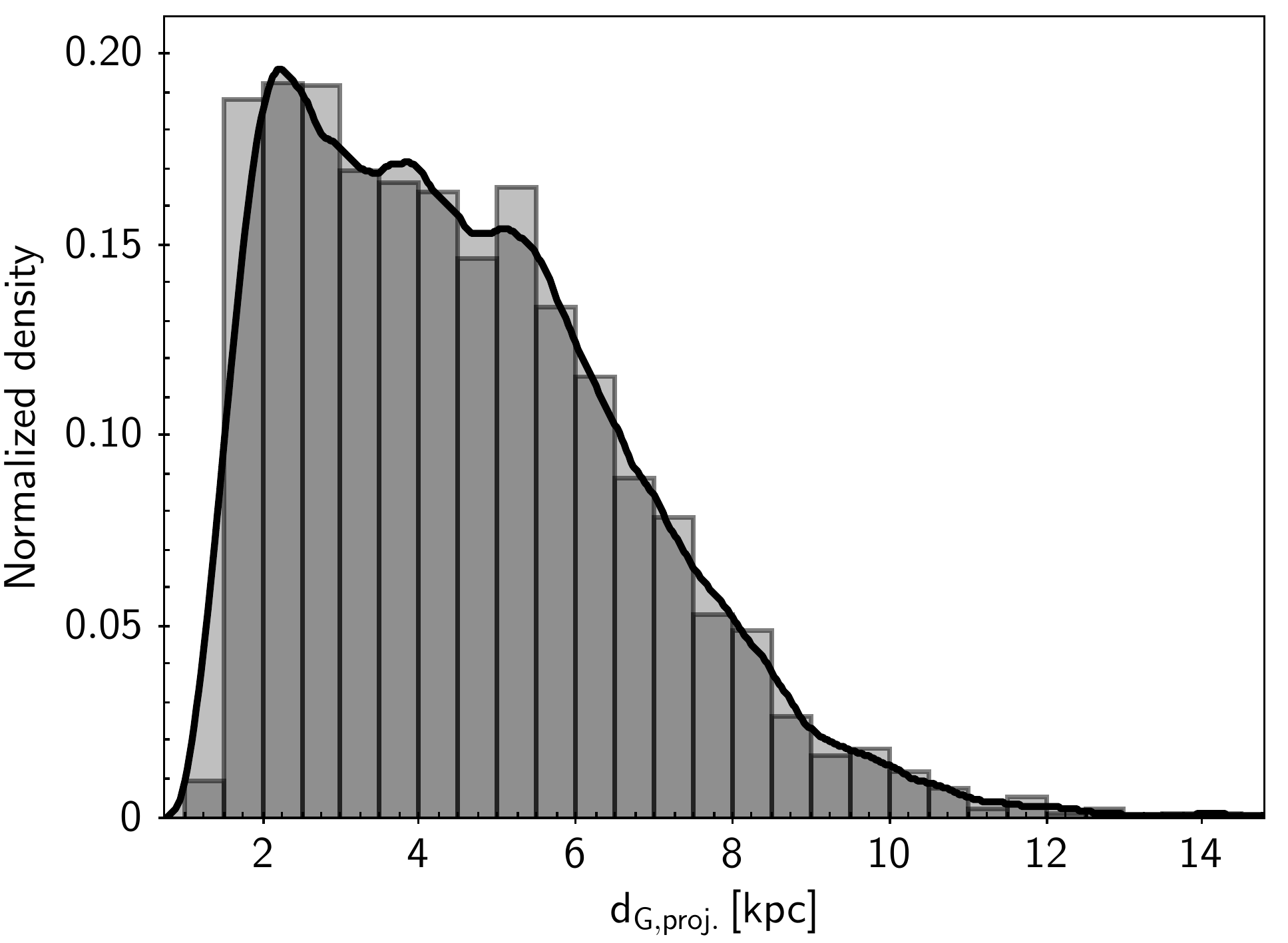}
\caption{Histogram {\em (gray bars)} and kernel density estimate {\em (black curve)} of the Galactocentric cylindrical distances of the RRLs. \label{fig:dghist}}
\end{figure}

The marginalized Galactocentric distance distribution also carries substantial selection biases (Fig.~\ref{fig:dists}, bottom right panel). 
An obvious feature is the slightly decreased number density at small $d_{G,proj.}$ and small $Z$, which is caused by very high extinction at low-latitude sight-lines close to the bulge ({\em cf.} Fig.~\ref{fig:spacedist}, middle panel). 
Another prominent feature is the smaller density of stars both far from the Galactic plane and far from the Galactic Center, but an important caveat is that this is {\em not} an intrinsic physical property of the distribution, but another selection effect due to the interplay between the conical geometry of the sample and the limiting magnitude: high-$Z$ stars are present only at large heliocentric distances, where completeness is lower due to the stars' fainter magnitudes, on average ({\em cf.} Fig.~\ref{fig:dists}, bottom left panel).

The bottom right panel of Fig.~\ref{fig:dists} shows the strong correlation between an object's location on the $(d_G,Z)$ plane and heliocentric distance $d_h$, arising from a combination of all the factors discussed above. In Sect.~\ref{subsec:mdf}, we will analyze the metallicity distribution of the objects in subsamples drawn from different marginalized Galactocentric distance ranges. Despite the strong selection effects in $d_h$, this is meaningful under the assumption that the [Fe/H] distribution of RRLs is circularly symmetric around the Galactic Center and along the Galactic plane over large spatial scales.

Finally, Fig.~\ref{fig:dghist} shows the histogram of the Galactocentric cylindrical distances of the RRL stars. The strong change in the derivative of the density at around 5.5--6\,kpc arises from the limited longitude range of the VVV, in combination with the survey's faint and bright limiting magnitudes. This is because the contribution of objects lying beyond 5.5--6\,kpc from the Galactic Center becomes systematically smaller for regions both at the far side of the disk and in the solar neighborhood, due to incompleteness.

In summary, due to the geometry of the surveyed volume, limiting magnitudes, and the selection effects due mainly to extinction, the current sample does not allow us to trace either the Galactocentric density profile of the RRL stars at high Galactocentric radii or their vertical density profile. These would require a more extended survey area in both longitude and latitude, and possibly a higher detection efficiency toward fainter magnitudes. We note that the former requirements will be met by both the OGLE-IV \citep{2015AcA....65....1U} and the VVV Extended (VVVX, \citealt{2018ASSP...51...63M}) surveys. In the following, due to the data limitations discussed above, we will constrain our analysis to spatial divisions made only in the marginalized projected (i.e., cylindrical) Galactocentric distance distribution of our stellar sample.

\subsection{Metallicity distribution function\label{subsec:mdf}}

\begin{figure}[]
\includegraphics[width=0.48\textwidth]{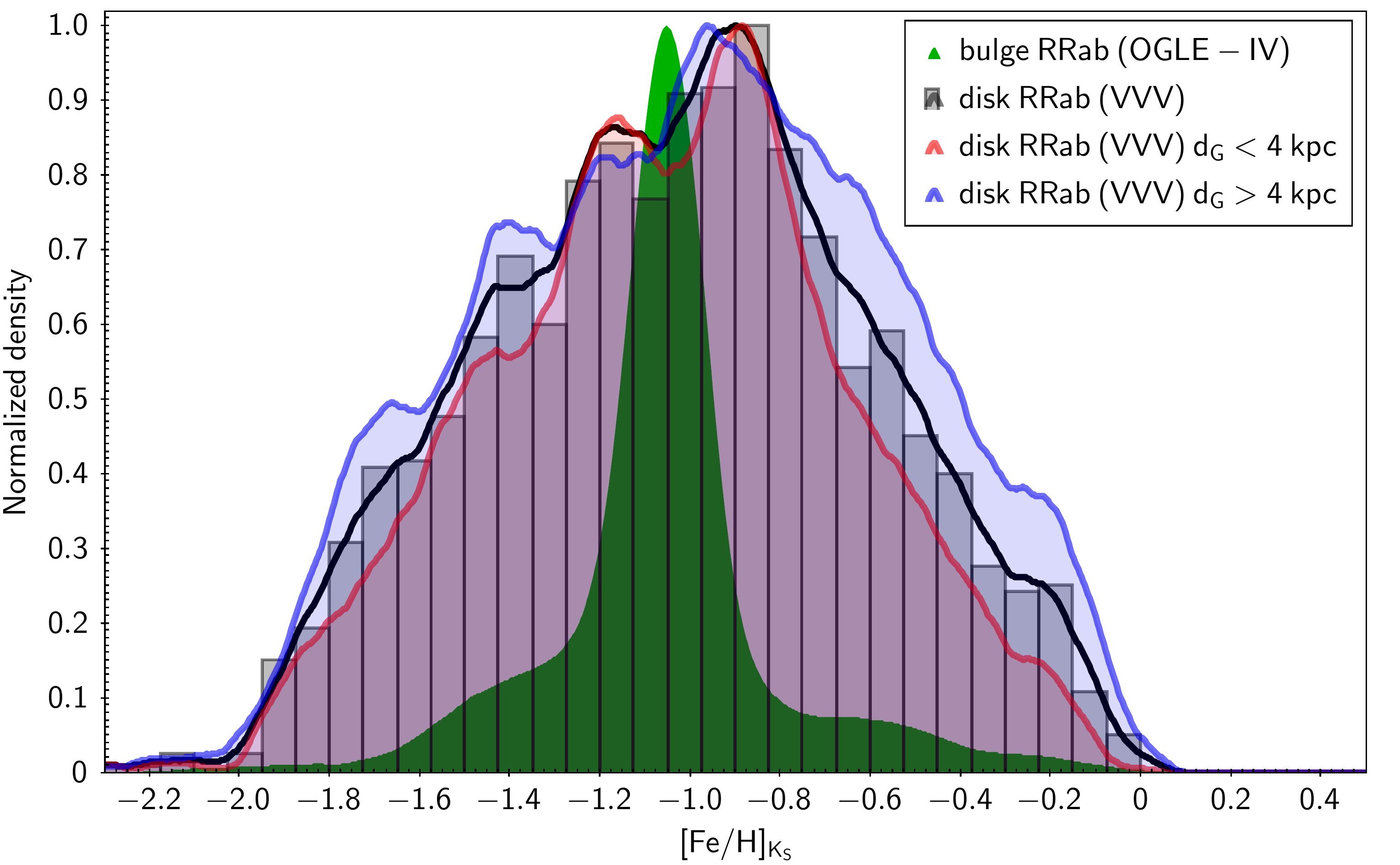}
\caption{MDFs of the RRLs stars, normalized to their maximum value. The gray histogram shows the [Fe/H] values of our full sample derived from the $K_s$ light curve parameters, while the black curve is a KDE. The red and blue curves show KDEs of the [Fe/H] of subsamples lying within and beyond a threshold of 4\,kpc Galactocentric radius, respectively. The green shaded area shows the KDE of the [Fe/H] of the 6193 bulge RRab stars from \cite{hajdu}, derived from their $I$-band light curve parameters \citep[][his Eq.~(3)]{2005AcA....55...59S}.\label{fig:mdf_rg}}
\end{figure}

The MDF of the RRLs estimated from their $K_s$-band light curve parameters (see Sect.~\ref{sec:feh}), after removing the poorest $\sim$10 \% of the data in terms of light curve fitting accuracy, is shown in Fig.~\ref{fig:mdf_rg} ({\em black curve}). We show it in direct comparison with the MDF of the same high-quality subset of OGLE-IV bulge RRab stars \citep{2015ApJ...811..113P} that were used as training data for the $K_s$-band metallicity predictor \citep[6193 objects,][see Sect.~\ref{sec:feh}]{hajdu}. The striking difference between the two MDFs is immediately evident, namely the MDF of the VVV disk RRL sample has much lower kurtosis, in contrast with the sharply peaked, narrow MDF of bulge RRab stars. At the same time, both distributions show clear signs of multi-modality.

We split our sample from the VVV disk field into two subsamples based on the objects' projected (cylindrical) Galactocentric distances, using a threshold of $d_G=4$\,kpc, which roughly halves our sample. The corresponding MDFs are shown in Fig.~\ref{fig:mdf_rg} with different colors. While the MDFs of both samples show the same phenomenological differences with respect to the MDF of the bulge RRLs, there is a relative excess of metal-rich objects (i.e., with [Fe/H]$\gtrsim -0.8$) farther away from the Galactic Center compared to the inner part of the disk, showing signs of two distinct metal-rich modes. The overwhelming majority of the RRLs in the OGLE-IV bulge sample, on the other hand, are located at $d_G<4$\,kpc and at higher Galactic latitudes, i.e., most of them reside within the bulge volume, with only a few percent of the objects lying in the foreground Galactic disk \citep{2015ApJ...811..113P}. In spite of the fact that they are sampled in rather different Galactic regions, signs of structural similarities can be observed between all MDFs, measured both toward the bulge, and toward the inner and outer parts of the disk, namely, that they exhibit a strong global maximum close to $-1$~dex, and that the modes of the distributions are centered at similar [Fe/H] values.

\begin{figure*}
\gridline{\fig{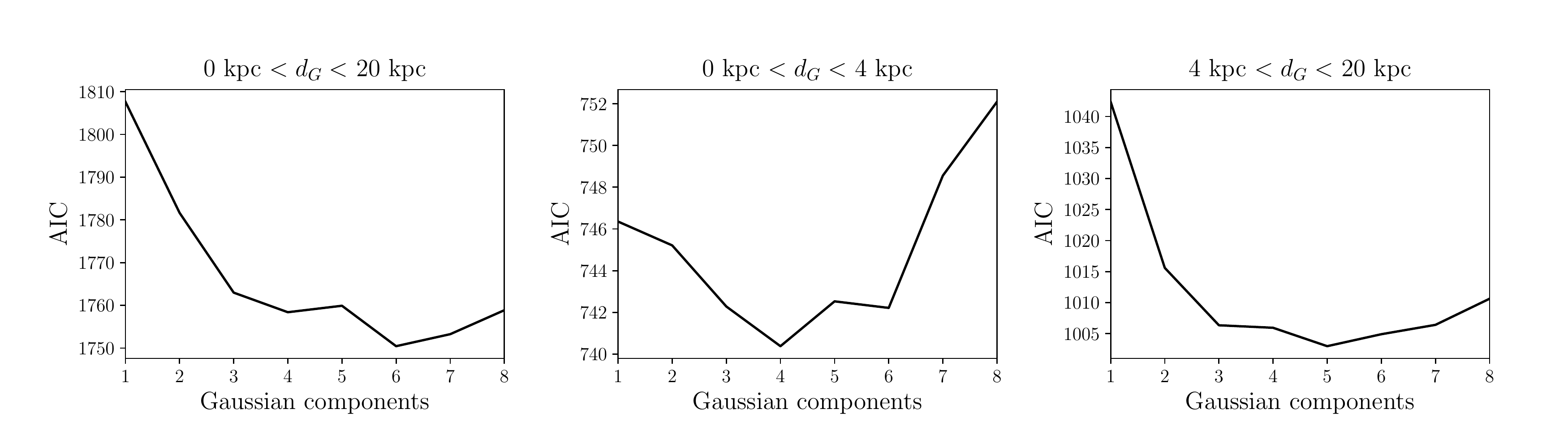}{0.9\textwidth}{}
          }
\gridline{\fig{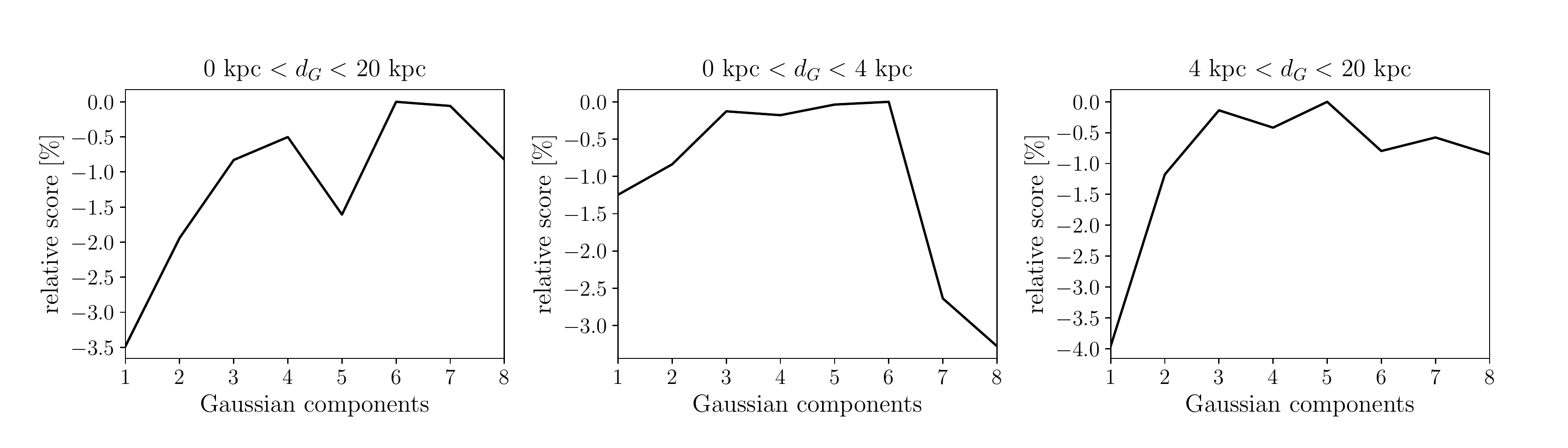}{0.9\textwidth}{}
          }
\caption{{\em Top row:} the AIC as a function of the number of components in the GMM of the RRL MDF, for the entire sample {\em (left)}, and for objects within {\em (middle)} and beyond a {\em (right)} 4~kpc Galactocentric radius. {\em Bottom row:} same as above, but showing the relative CV scores on the vertical axes. \label{fig:complexity}}
\end{figure*}

\begin{figure*}
\centering
\gridline{\fig{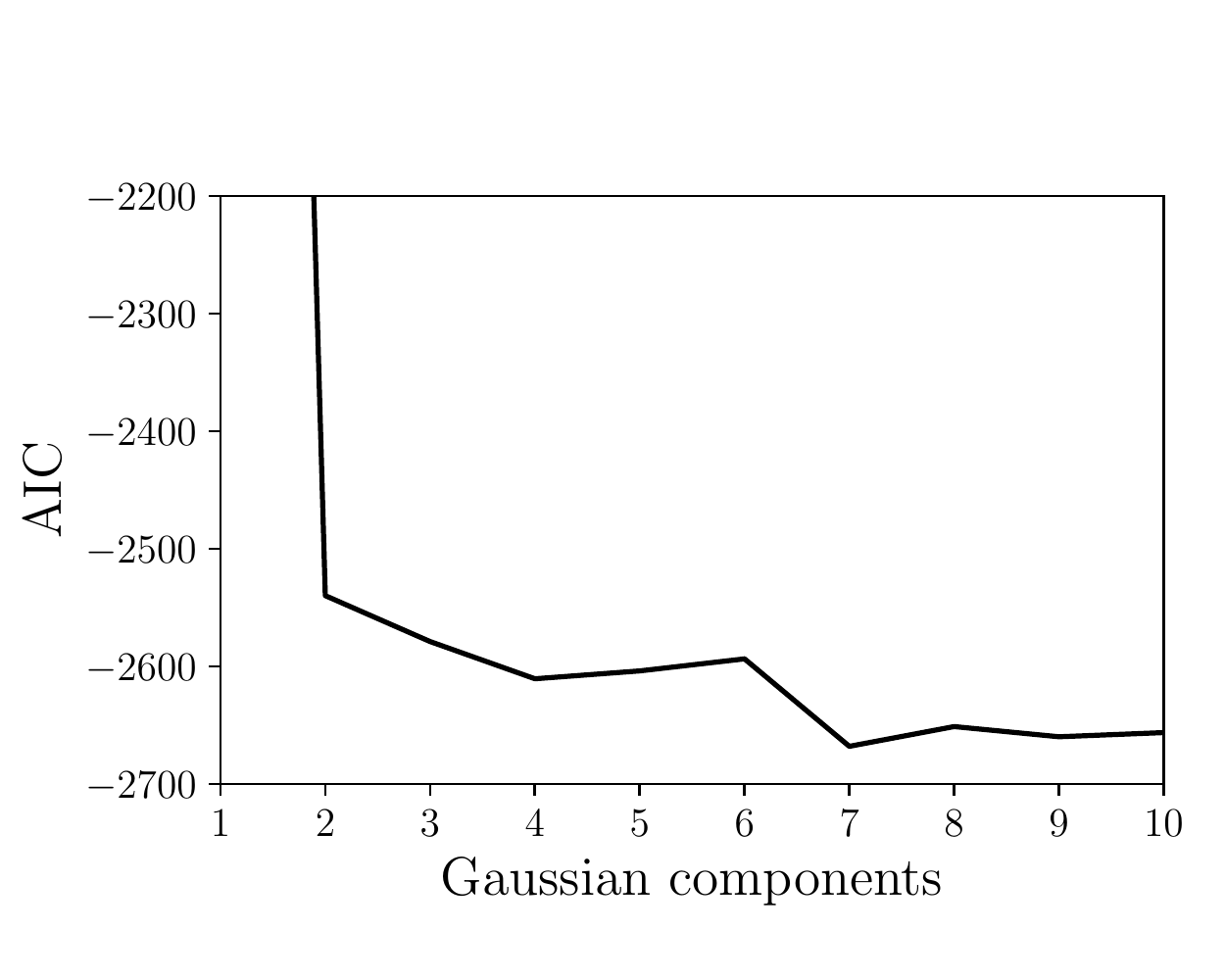}{0.35\textwidth}{} \fig{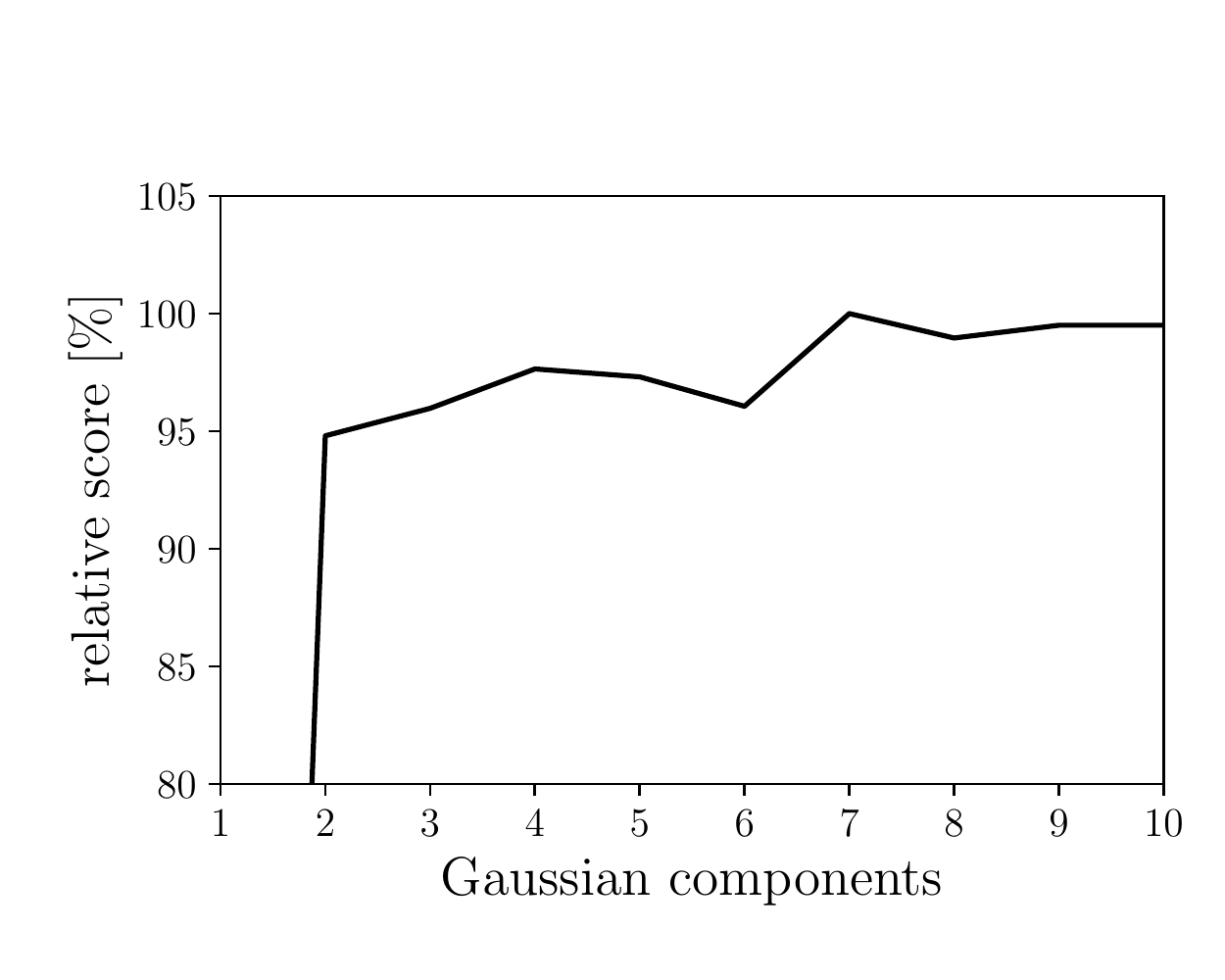}{0.35\textwidth}{}}
\caption{AIC ({\em left}) and relative CV scores ({\em right}) plotted against the number of components in the Gaussian mixture model of the bulge RRL MDF. \label{fig:complexity_bulge}}
\end{figure*}

In order to enable a quantitative comparison of the RRL MDFs observed toward the bulge and the southern disk with each other, and with independent results later in Sect.~\ref{subsec:mdf_comp}, we used a mixture of Gaussian distribution functions as their common mathematical representation. The Gaussian mixture models (GMMs) were fitted to the data with the maximum likelihood method, via the expectation maximization algorithm \cite[][see their Chapter 4.4 and references therein]{2014sdmm.book.....I} implemented in the scikit-learn software package \citep{2011JMLR....12.2825}. We optimized the model complexity (i.e., the number of Gaussian components) by two different approaches.

First, we took a classical information theory approach, the Akaike information criterion \citep[AIC,][]{1974IEEE..19....6}. The AIC measures relative information loss in an asymptotic approximation, and has the form:

\begin{equation}
{\rm AIC} = 2k_{\rm(M)} - 2\ln{\hat{L}{\rm (M)}} + \frac{2k_{\rm M}(k_{\rm M}+1)}{N-k_{\rm M}-1},\label{eq:aic}
\end{equation}

\noindent where $k$ is the number of parameters, $\hat{L}(M)$ is the maximum value of the likelihood function of model M, and $N$ is the length of the data (in our case, the number of [Fe/H] values). In Eq.~\ref{eq:aic}, the first term penalizes model complexity against the goodness of fit measured through the log-likelihood in the second term, while the third term is a correction for the finite number of data points \citep[see, e.g., ][]{2007MNRAS.377L..74L}. The best model representation is identified by minimizing the value of AIC as a function of $k$ for a certain family of models. 

\begin{figure*}
\centering
\includegraphics[width=0.85\textwidth]{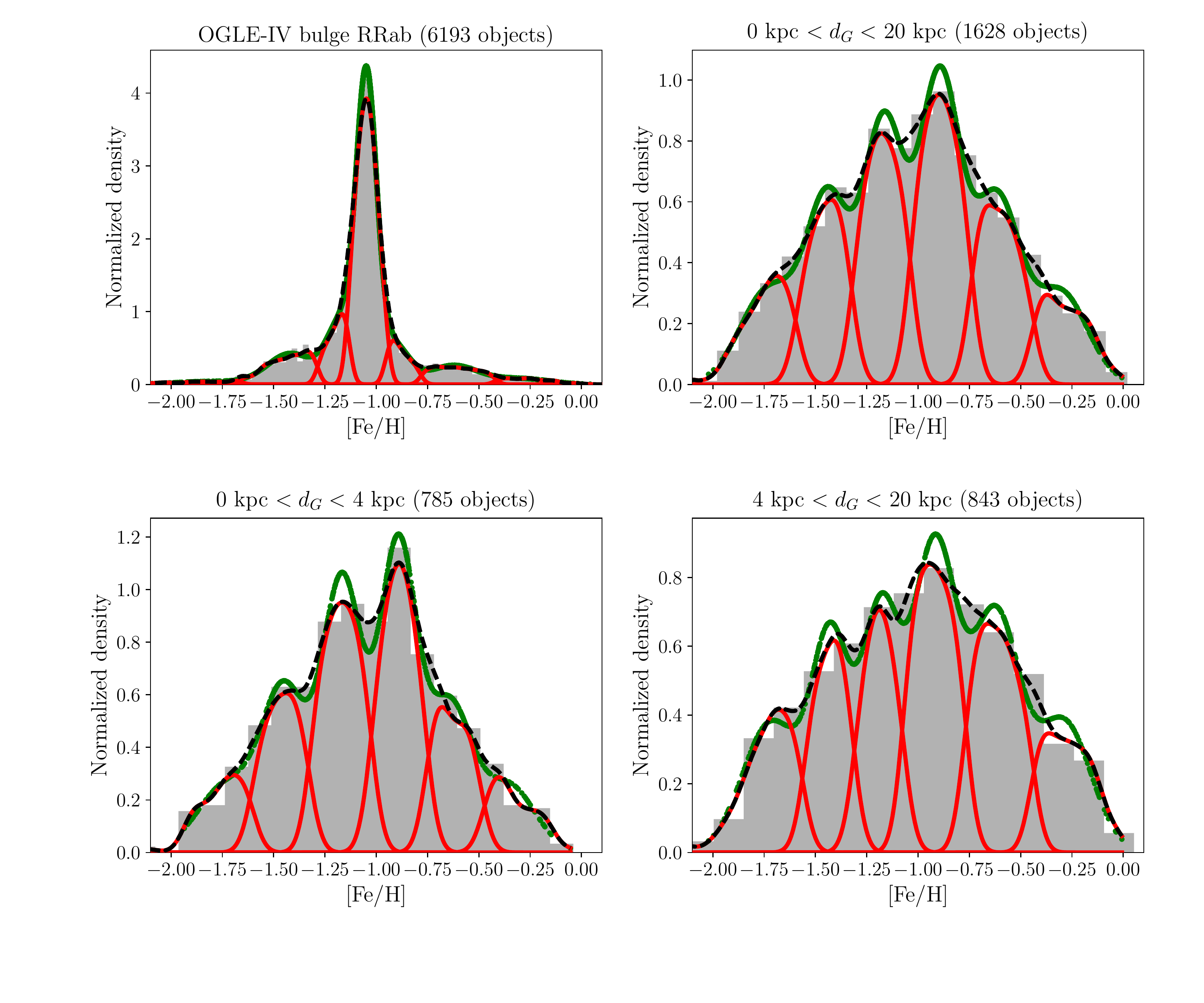}
\caption{MDFs of the OGLE-IV RRab stars in the bulge {\em (top left)} compared to the RRL MDFs obtained in this study, for samples taken from different ranges of Galatocentric distance (as marked in the figure headers). The {\em gray bars} show the histogram of photometric [Fe/H] estimates, while the {\em black curves} denote their kernel density estimates (KDEs), both normalized to their integrals. The {\em green lines} show the GMMs fitted to the data, and {\em red curves} show the KDEs of the objects in the distinct GMM components. \label{fig:gmm}}
\end{figure*}

The optimal numbers of Gaussian components in the GMMs were also assessed by a standard $K$-fold cross-validation (CV) procedure. In this procedure, the data are randomly divided into $K$ folds, and each fold is used as a validation data set. The prediction score (in our case, the log-likelihood) is computed for each fold by training the model with the data in the other $K-1$ folds. This is done $K$ times, and the resulting average score is used as a performance metric of the model. This procedure is repeated for each model complexity (in our case, each value of $k$), and the model that yields the highest CV score is accepted as optimal. We performed this analysis with $K=20$, in order to ensure that the model parameters derived in each fold do not become biased due to holding out too much data for the validation set.

Figure~\ref{fig:complexity} shows the value of AIC and the relative CV score against the number of components $k$ in the GMMs fitted to the [Fe/H] values of our sample of RRLs observed toward the southern Galactic disk. As explained above, the optimal model corresponds to the minimum AIC value and the maximum CV score. The optimal number of components is found to be 6 by both methods when objects from our full Galactocentric distance range are considered. We note that for subsamples of the data, the optimal number of GMM components varies between 4 and 6 due to the combined effect of the varying weights of the MDF modes with $d_G$, and the noisiness and limited sizes of the subsamples. The MDF is phenomenologically similar within all Galactocentric distance ranges and it does not become truncated with respect to the MDF of the full sample, therefore we fitted a six-component GMM in all subsamples.

We performed similar AIC and twentyfold CV analyses to determine the optimal GMM complexity for the OGLE-IV bulge RRL sample. The results are shown in Fig.~\ref{fig:complexity_bulge}. In both cases, a seven-component GMM was found to be optimal. We note that the weak and long metal-poor and metal-rich tails of the bulge RRL MDF give rise to rather shallow extrema in both kinds of complexity analyses, thus not all seven components are readily identifiable by eye in Fig.~\ref{fig:mdf_rg}.

Figure~\ref{fig:gmm} visualizes the results of the GMMs for the same data sets as in Fig.~\ref{fig:mdf_rg}. We show histograms of the [Fe/H] values, overlaid by the mixture models evaluated at the data points (in green), and kernel density estimates (in red) of the different model components using Gaussian kernels, with optimal kernel sizes evaluated via twentyfold CV. We performed Monte Carlo (MC) simulations to determine the errors in the GMM parameters of the MDFs using Gaussian random noise. In every MC realization, each [Fe/H] value was modified by being drawn from a Gaussian distribution centered at the original [Fe/H] value, and with its standard deviation made equal to the [Fe/H] prediction error of 0.2~dex and 0.14~dex in case of the VVV disk and OGLE-IV bulge samples, respectively (see Sect.~\ref{sec:feh}). The errors in the parameters of the GMM components were determined from 1000 realizations. Tables~\ref{tab:gmm} and \ref{tab:gmm_bulge} present the parameters of the GMM components obtained for the various RRL subsamples observed toward the southern Galactic disk (VVV fields) and the Galactic bulge (OGLE fields), together with their errors, labeled with `v' and `o', respectively, and numbered with decreasing metallicity. We note that although the errors in the weights are relatively large, they are not independent for a given subsample.

\begin{longrotatetable}
\begin{deluxetable*}{c|ccc|ccc|ccc}
\tablecaption{Parameters of the GMM Components of the RRL MDFs\tablenotemark{1} Observed toward the Southern Galactic Disk. \label{tab:gmm}}
\tablewidth{0pt}
\tablehead{
  & \multicolumn{3}{c|}{All $d_G$} & \multicolumn{3}{c|}{$d_G < 4$~kpc} & \multicolumn{3}{c}{$d_G > 4$~kpc} \\
  Component &  Mean [Fe/H] & $\sigma$ & Weight & Mean [Fe/H] & $\sigma$ & Weight & Mean [Fe/H] & $\sigma$ & Weight
}
\startdata
   v1    &  $-0.30 \pm 0.05 $   & $0.130 \pm 0.007 $ & $ 0.10 \pm 0.03$ &      $-0.36  \pm 0.10$ & $0.129 \pm 0.010$  & $0.08 \pm 0.03$ &       $-0.29 \pm 0.08$ & $ 0.132 \pm 0.009$ &  $0.12 \pm 0.04$     \\
   v2    &  $-0.62 \pm 0.06 $   & $0.115 \pm 0.002 $ & $ 0.17 \pm 0.03$ &      $-0.64  \pm 0.09$ & $0.107 \pm 0.003$  & $0.15 \pm 0.04$ &       $-0.62 \pm 0.09$ & $ 0.120 \pm 0.004$ &  $0.20 \pm 0.04$     \\
   v3    &  $-0.90 \pm 0.07 $   & $0.101 \pm 0.002 $ & $ 0.25 \pm 0.02$ &      $-0.89  \pm 0.09$ & $0.089 \pm 0.002$  & $0.26 \pm 0.03$ &       $-0.92 \pm 0.09$ & $ 0.108 \pm 0.003$ &  $0.24 \pm 0.03$     \\
   v4    &  $-1.17 \pm 0.07 $   & $0.100 \pm 0.002 $ & $ 0.21 \pm 0.02$ &      $-1.16  \pm 0.09$ & $0.099 \pm 0.002$  & $0.26 \pm 0.03$ &       $-1.18 \pm 0.10$ & $ 0.093 \pm 0.003$ &  $0.16 \pm 0.03$     \\
   v5    &  $-1.44 \pm 0.07 $   & $0.103 \pm 0.003 $ & $ 0.15 \pm 0.03$ &      $-1.45  \pm 0.10$ & $0.109 \pm 0.004$  & $0.16 \pm 0.04$ &       $-1.43 \pm 0.10$ & $ 0.092 \pm 0.004$ &  $0.14 \pm 0.03$     \\
   v6    &  $-1.72 \pm 0.06 $   & $0.141 \pm 0.007 $ & $ 0.11 \pm 0.03$ &      $-1.72  \pm 0.12$ & $0.147 \pm 0.011$  & $0.09 \pm 0.04$ &        $-1.71 \pm 0.11$ & $ 0.140 \pm 0.010$ &  $0.13 \pm 0.04$     \\
\enddata
\tablenotetext{1}{[Fe/H] values are on the \cite{1995AcA....45..653J} metallicity scale (see Sect.~\ref{sec:feh}).}
\end{deluxetable*}
\end{longrotatetable}

\begin{deluxetable*}{c|ccc}
\tablecaption{Parameters of the GMM Components of the RRL MDF\tablenotemark{1} Observed toward the Galactic Bulge. \label{tab:gmm_bulge}}
\tablewidth{0pt}
\tablehead{
  Component &  Mean [Fe/H] & $\sigma$ & Weight 
  }
\startdata
   o1    &  $-0.30 \pm  0.05$   & $0.148 \pm  0.005$ & $0.03  \pm 0.01$   \\
   o2    &  $-0.63 \pm  0.05$   & $0.120 \pm  0.002$ & $0.08  \pm 0.01$   \\
   o3    &  $-0.92 \pm  0.03$   & $0.078 \pm  0.001$ & $0.11  \pm 0.03$   \\
   o4    &  $-1.05 \pm  0.03$   & $0.049 \pm  0.001$ & $0.49  \pm 0.01$   \\
   o5    &  $-1.17 \pm  0.03$   & $0.074 \pm  0.001$ & $0.16  \pm 0.03$   \\
   o6    &  $-1.42 \pm  0.03$   & $0.111 \pm  0.001$ & $0.12  \pm 0.02$    \\
   o7    &  $-1.79 \pm  0.05$   & $0.194 \pm  0.006$ & $0.02  \pm 0.01$   \\
\enddata
\tablenotetext{1}{[Fe/H] values are on the \cite{1995AcA....45..653J} metallicity scale (see Sect.~\ref{sec:feh}).}
\end{deluxetable*}

\subsection{Spatial variations in the metallicity distribution}\label{subsec:spatial_mdf}

\begin{figure*}[]
\centering
\gridline{\fig{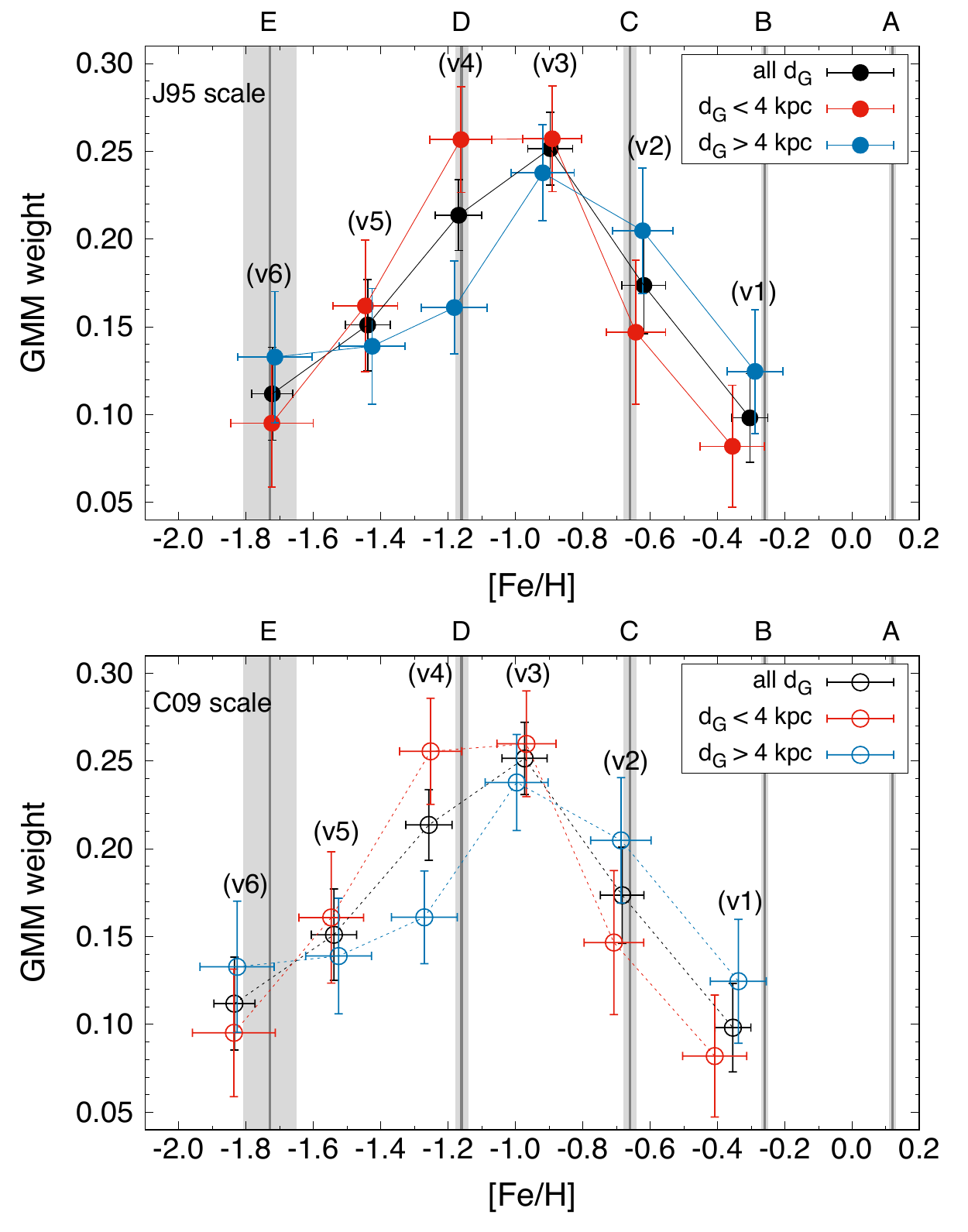}{0.45\textwidth}{}
              \fig{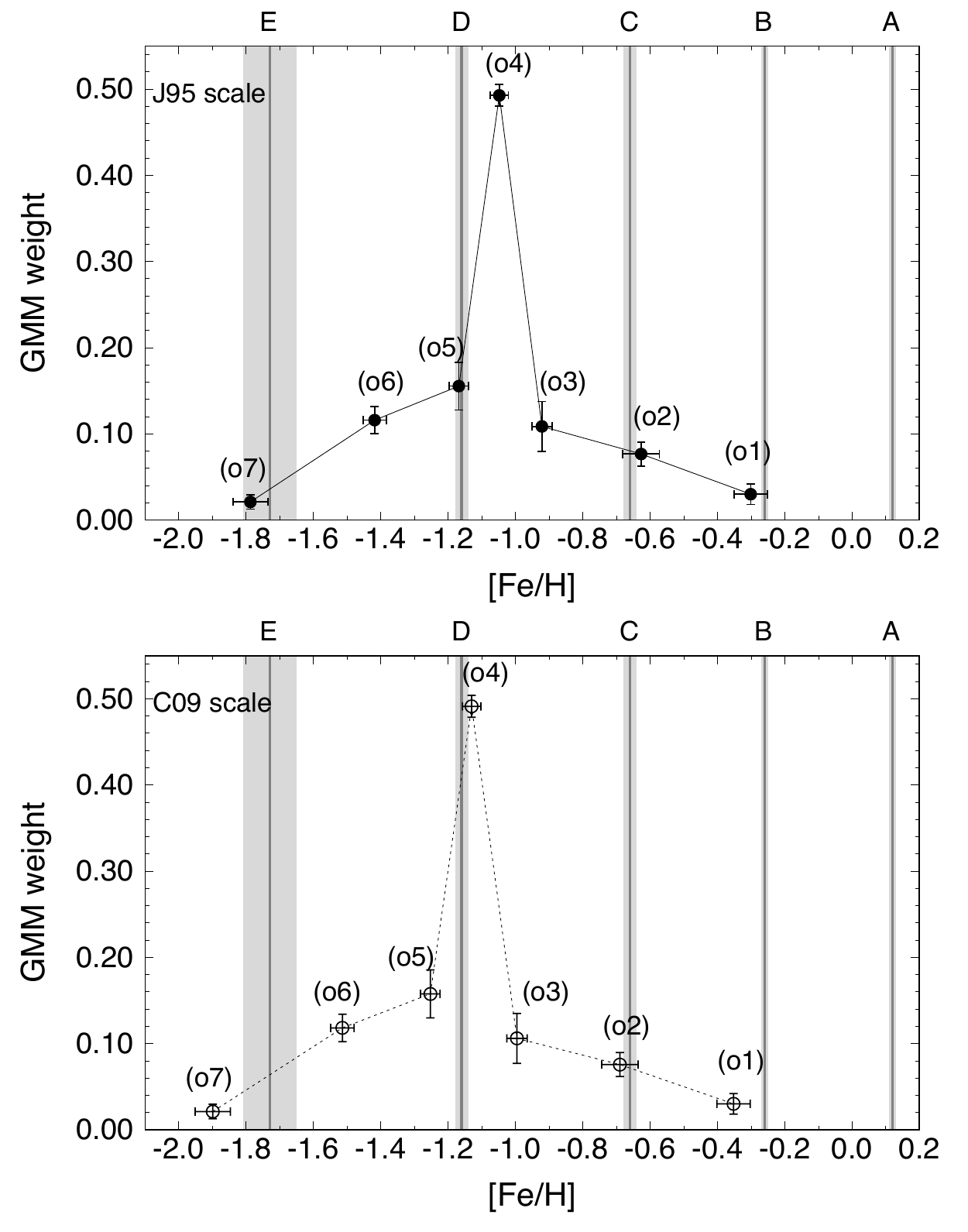}{0.45\textwidth}{}
}

\caption{The weights and means of the GMM components of the RRL MDFs observed toward the VVV disk fields ({\em left}) and the OGLE-IV bulge fields ({\em right}). In the left panels, different colors show model parameters obtained for objects lying in different ranges of Galactocentric distances ($d_G$) as indicated in the figure keys. The error bars represent statistical errors derived by MC simulations (see text). The {\em top panels} show the results on the \cite{1995AcA....45..653J} metallicity scale (on which the [Fe/H] prediction is performed), while the {\em bottom panels} shows the model parameters derived from the [Fe/H] distributions transformed to the C09 metallicity scale, using the transformation equation of \cite{hajdu}.
The {\em gray vertical lines} mark the values of the GMM component means identified in the MDF of the ARGOS bulge sample at $b=-5^\circ$, $l=\pm15^\circ$ and $d_G\leq3.5$~kpc \citep[][see their Table 3; the values are provided on their own metallicity scale]{2013MNRAS.430..836N}. Their statistical errors are indicated by vertical shaded areas, while their identifiers are shown at the top of the panels, following the notation of \cite{2013MNRAS.430..836N}.
\label{fig:gmm_comps}}
\end{figure*}

The top left and top right panels of Fig.~\ref{fig:gmm_comps} show the weights of the resulting GMM components against their means for the RRL MDFs observed toward the disk and the bulge, respectively. The means of the former are very stable against $d_G$, well within their errors, which is remarkable, considering that the two subsamples are from distinct $d_G$ ranges (i.e., they have no common objects). In all three MDFs obtained for the VVV disk fields, the global maxima fall to $[{\rm Fe/H}] \simeq -0.9$, and the corresponding mode has the largest weight at all values of $d_G$. The relative weights of both metal-rich components with $[{\rm Fe/H}] > -1$ become larger with increasing distance from the Galactic Center, with respect to the weights of the more metal-poor modes. This is consistent with a small positive Galactocentric metallicity gradient along the disk. Likewise, the weights of the metal-poor ($[{\rm Fe/H}] < -1$) modes systematically decrease with increasing $d_G$, except for the weakest metal-poor mode at $[{\rm Fe/H}] \simeq -1.70$. This trend consistently extrapolates toward the bulge: inside the bulge volume (i.e., in case of the OGLE bulge RRLs), the metal-rich modes of the MDF dissolve into a weak tail, while the metal-poor tail remains relatively more significant.

Although only a small fraction of RRLs in the OGLE bulge fields contribute to the metal-poor and metal-rich wings of the MDF, the means of the corresponding GMM modes match those in the VVV disk field MDFs with very high precision ({\em cf.}~Tables~\ref{tab:gmm} and \ref{tab:gmm_bulge}), capturing a significant structural similarity between the distributions. Namely, the positions (i.e., the means) of components v1, v2, v5, and v6 match components o1, o2, o6, and o7, well within the statistical uncertainties. Components v3 and v4 can be associated with a mixture of components o3, o4, and o5. Although nearly half of the RRLs in the OGLE bulge sample correspond to component o4, it is the narrowest one with $\sigma=0.05$~dex. Importantly, when associating the GMM modes between the different MDFs around $-1$~dex, we must take into consideration the following caveats: (i) the MDF of the OGLE RRLs toward the bulge has a higher precision ($\sim0.14$~dex), since it was derived from the $I$-band light curves; (ii) at the same time, components o3--o5 are heavily blended, perhaps due to the dominance of component o4 and/or the nonnormality of the underlying physical MDF components; and finally, (iii) the $K_s$-band [Fe/H] prediction may be slightly biased toward higher metallicities around $-1$~dex \citep[see][]{hajdu}. Consequently, more precise metallicity estimates are required to study the fine structure of the MDF around its global maximum.

\begin{figure}[]
\includegraphics[width=0.5\textwidth]{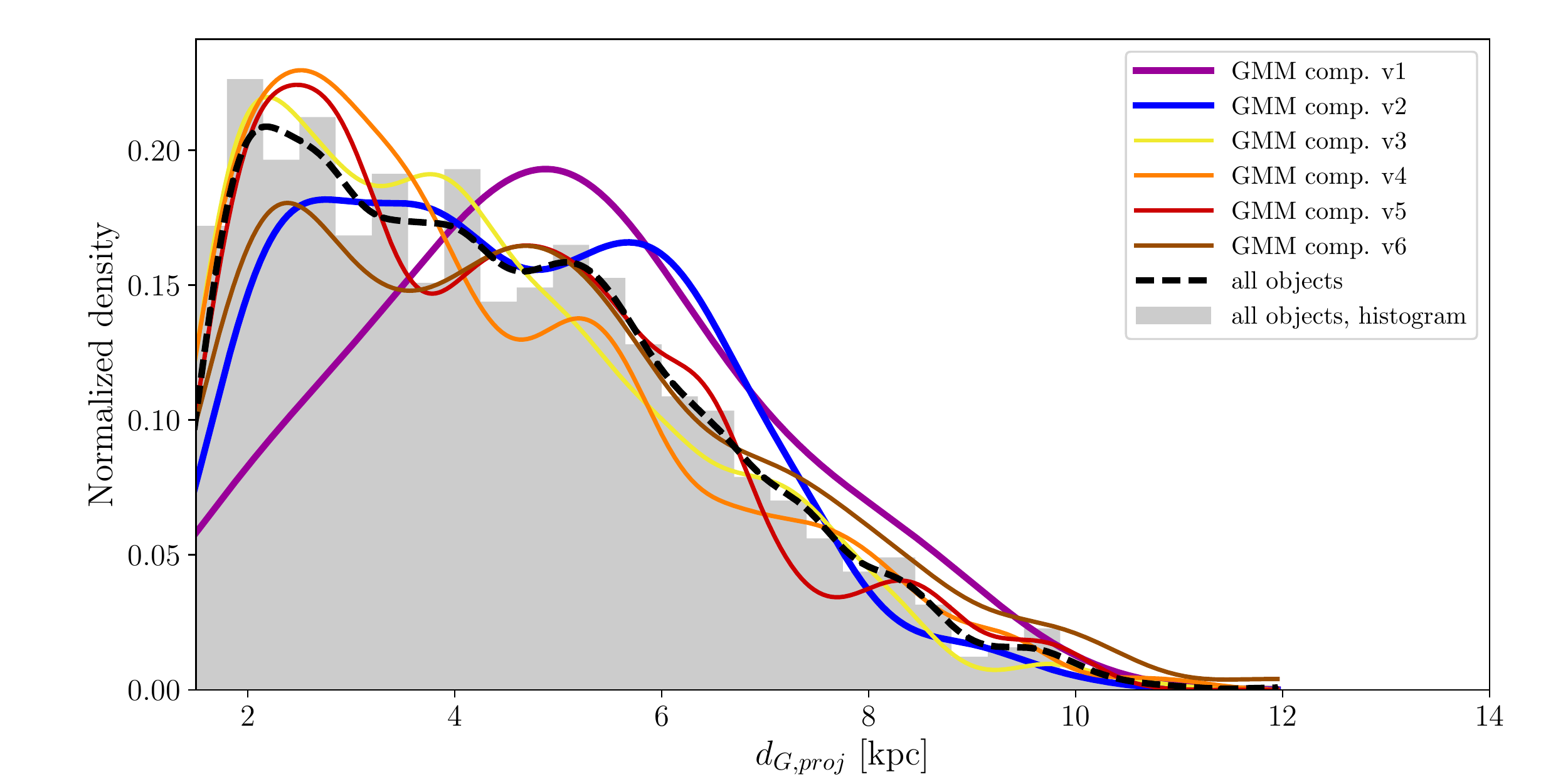}
\caption{Galactocentric cylindrical distance ($d_{G,proj.}$) distributions of the RRL subsamples from the VVV disk fields corresponding to the different components of their MDF. Curves in different colors show kernel density estimates of the various components as indicated in the figure key. The histogram of all stars in our analysis is shown in gray.\label{fig:gcdist_comp_kde}}
\end{figure}

\begin{figure}[]
\includegraphics[width=0.5\textwidth]{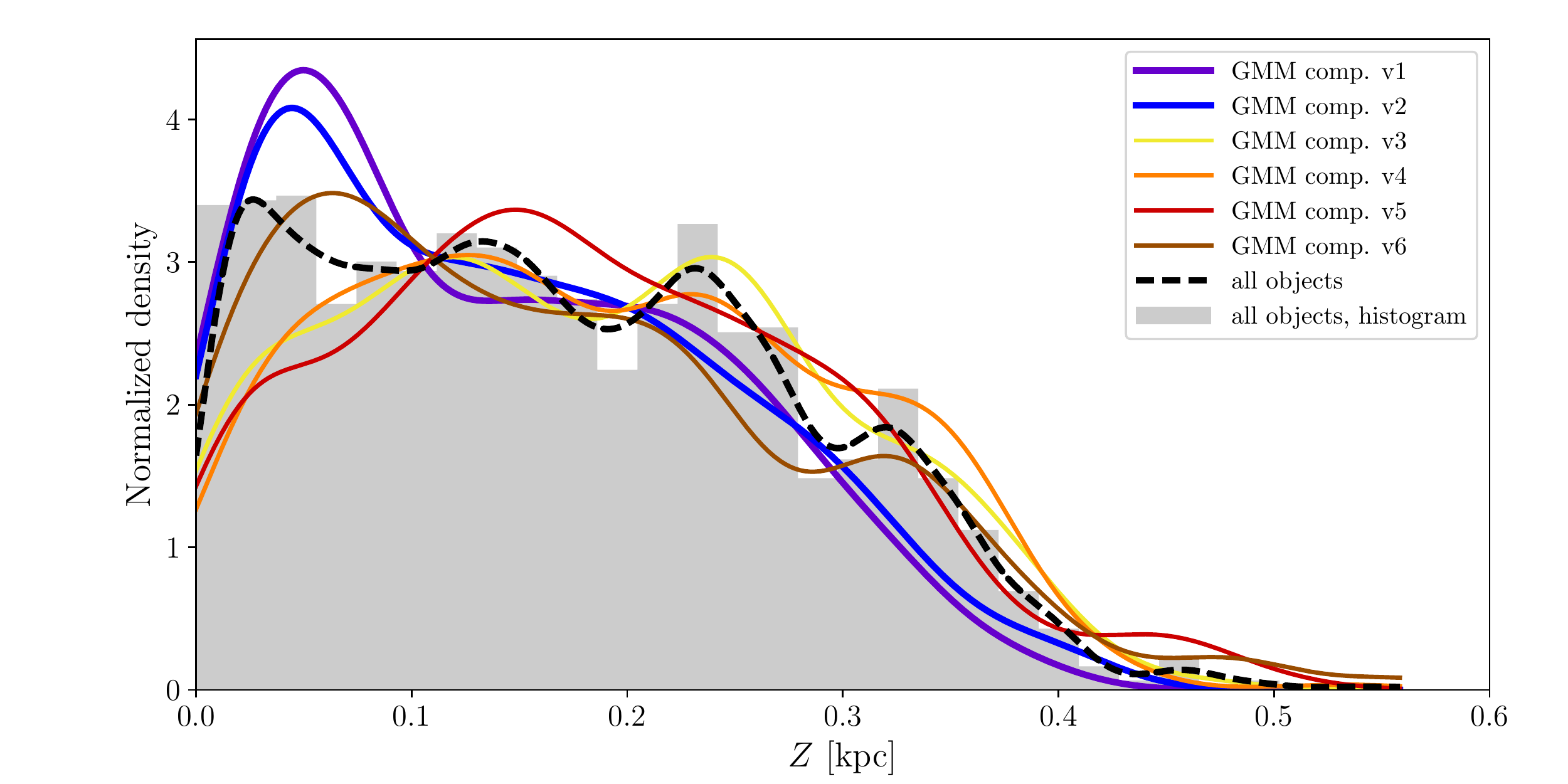}
\caption{Same as Fig.~\ref{fig:gcdist_comp_kde}, but showing the distributions of the distances from the Galactic plane ($Z$). \label{fig:zdist_comp_kde}}
\end{figure}

Finally, we note that for a very small number of stars, extremely low metallicities ($[{\rm Fe/H}]<-2$) were estimated from their $K_s$-band light curves. Owing to the lack of extensive training data, our predictive model becomes unreliable for $[{\rm Fe/H}] \lesssim -1.8$ \citep{hajdu}, therefore these estimates should be treated with due caution. At the same time, there should be a weak metal-poor part in the MDF, since the spectroscopically identified most metal-deficient RRL stars in the field have $[{\rm Fe/H}] \simeq -2.7$ \citep{2011ApJ...743L...1H}. This is reinforced by the visual inspection of the light curves of the RRLs with the lowest metallicities, i.e., they are clearly not spurious [Fe/H] estimates from bad-quality data. However, the limitations of our current [Fe/H] predictor do not allow us to further expound the properties of these objects, which are therefore not subjects of our main discussion.

We further investigate the dependence of the metallicity on the Galactocentric distance and the distance from the Galactic plane by splitting the data into subsets according to their MDF components, based on the probability density function of the fitted GMM. Figure~\ref{fig:gcdist_comp_kde} shows the Galactocentric distance distribution of the RRLs found in the VVV disk field, constituting the MDF components v1--v6 by means of kernel density estimates. A Gaussian kernel was used and the bandwidth (kernel size) was optimized for each data subset via 20-fold CV. While the metal-poor components v3--v6 follow well the distribution of the full sample, the relative positions of maximum density of components v1 and v2 are clearly offset, showing that the relative number density of stars with higher metallicities increases with Galactocentric distance, and conversely, metal-poor stars are concentrated toward the inner Galaxy. We interpret this as a positive mean metallicity gradient of the RRLs along the Galactic disk. The number density distribution along the perpendicular spatial dimension, i.e., as a function of the $Z$ distances from the Galactic plane, is shown in Fig.~\ref{fig:zdist_comp_kde}. The kernel density estimates were computed in the same way as for Fig.~\ref{fig:gcdist_comp_kde}. We can clearly observe a dependence on metallicity: stars that correspond to the metal-rich components v1 and v2 are more concentrated toward the plane than the rest of the sample.

\subsection{Comparison of the MDF to independent results}\label{subsec:mdf_comp}

In the following, we compare our results to a recent measurement of the MDFs in the southern disk and the inner Galaxy based on data from the ARGOS survey, analyzed by \cite{2013MNRAS.430..836N}. 
Their MDFs are based on spectroscopic observations of a large sample of stars consisting of mainly RC giants. 
Their measurements were carried out at a relatively high resolution of $R\simeq11000$ \citep[see][for details]{2013MNRAS.428.3660F}, resulting in individual [Fe/H] values that are precise to $\sim$0.09~dex. 
The footprint of their survey covered several fields toward the bulge, and also the southern disk (where their target density was significantly smaller).
They measured a multi-modal MDF, which was subjected to a Gaussian decomposition analysis similar to ours. 
It is instructive, therefore, to perform a quantitative, direct comparison of the results from the two analyses, which are based on entirely independent [Fe/H] measurements.

The relative fraction of metal-rich stars in the ARGOS sample is much larger compared to the RRab stars in our study. The ARGOS MDFs are dominated by objects with $[{\rm Fe/H}] > -0.5$, and contain a relatively minor fraction of metal-poor stars at $[{\rm Fe/H}] < -1$. However, the ARGOS and the RRL MDFs show a remarkable similarity as far as the positions of the distribution's modes, i.e., the means of the GMM components, are concerned. 

Figure~\ref{fig:gmm_comps} compares our results with the ARGOS MDF components, by visualizing the means of the \cite{2013MNRAS.430..836N} GMM model components labeled as A--E and their errors for stellar samples in the bulge at low Galactic latitudes. We note that the component means from their disk field are the same to within their uncertainties (see also Sect.~\ref{subsec:origins}), but their disk MDF consists of many fewer stars. There is a precise match between the positions of the two metal-rich components of the models, namely the ARGOS components B and C and the RRL components v1, v2, and o1, o2, respectively. The metal-poor ARGOS components D and E match our RRL MDF components v4, v6 and o5, o7 with similarly high precision. These matches are especially remarkable if we consider the very different weights with which the metal-poor components are present in the three data sets, and the fact that the matches are well within the combined statistical errors. For most of the ARGOS subsamples, component B is the most prominent, with a weight varying between 0.4--0.5, while it gives a $\sim$30\% and $\sim$90\% smaller relative contributions to our VVV disk and OGLE bulge RRL MDFs, respectively. In spite of this, the agreement is better than 0.05~dex in case of the OGLE bulge RRLs, and the full VVV disk RRL sample and outer disk subsample. The position of component C, which also has a large weight in the ARGOS MDFs, has a similarly good match with the RRL data. Importantly, the most metal-rich ARGOS component A is not present in either the VVV disk, or the OGLE bulge RRL MDFs.

Despite the fact that component E is weak in both the ARGOS and the RRL MDFs, the position of this mode also matches well in the three data sets, while our components v5 and the corresponding o6 do not have a counterpart in the ARGOS MDF. The main difference in the modes of the two MDF decompositions is observed for metallicities close to ${\rm [Fe/H]} \approx -1$~dex. The strongest mode in the VVV disk RRL MDF (component v3) falls between the ARGOS components C and D, peaking at $[{\rm Fe/H}] \simeq -0.9$, and likely corresponds to the blend of components o3+o4; it is identified as the counterpart of the center of the bulge RRL MDF.

We emphasize that the methods used for the determination of [Fe/H] in the two studies are completely different, hence a systematic offset of $\sim$0.1~dex is not unexpected between the ARGOS and the J95 metallicity scales. Unfortunately, a transformation between the ARGOS and other, commonly used metallicity scales is not established. For completeness, we transformed our metallicities from the J95 scale to the C09 scale using the transformation formula provided by \cite{hajdu}, and show the parameters of the corresponding GMMs in the lower panel of Fig.~\ref{fig:gmm_comps}. The resulting MDFs are very similar, with an increasing systematic offset in the component means toward smaller metallicities. The match between the ARGOS and the RRL GMM modes is preserved for the VVV disk MDF, remaining within the combined statistical errors, while component o4 in the OGLE bulge RRL MDF becomes shifted to nearly match the ARGOS component D.

\section{Discussion} \label{sec:discussion}

\subsection{The ages of RR~Lyrae stars}

In Sect.~\ref{subsec:spatial_mdf}, we compared the MDF of the RRL stars, resulting from our study, to the MDF of RC giants based on the ARGOS survey. RC stars span an age range of 1--10~Gyr, and correspondingly a wide range of masses (and metallicities) \citep{2001MNRAS.323..109G,2016ARA&A..54...95G}. In the bulge though, we may assume that most of them are close to the older age range though \citep[e.g.,][]{1995Natur.377..701O,2010ApJ...725L..19B}. Their frequency peaks at ages of a few Gyr in galaxies with extended star formation histories \citep{2016ARA&A..54...95G}.  
In contrast, the advantage of RRLs compared to other tracer objects, such as RC stars, is that the ages of RRLs are better constrained, and they represent true stellar fossils. \cite{2008AJ....135.1106G} studied the age of the globular cluster NGC\,121 based on very deep {\em Hubble Space Telescope} observations, and found ages ranging between $10.5\pm0.5$ and $11.8\pm0.5$~Gyr, depending on the stellar evolution models and the photometric diagnostics used. As pointed out by \citet{2009IAUS..258..209C}, this is the youngest globular cluster known to contain RRLs, and is 2 to 3~Gyr younger than the oldest globular clusters in the Milky Way, the Large Magellanic Cloud, and the Fornax and Sagittarius dwarf galaxies, suggesting that 10~Gyr is a good assumption for the (empirical) lower age limit for RRL stars. Due to the systematic age difference between the RC and RRL stars, the differences in the weights of their MDF components must be, at least to first order, resulting from the age differences of the underlying stellar populations.  

\subsection{The RR~Lyrae production efficiency}\label{subsec:rpef}

In attributing the differences between the RRL and ARGOS MDFs to the differences in the age ranges of the tracer objects used, we must take into account the production efficiency of the two types of stars, as a function of the various stellar parameters, other than the age. The MDF that we observe using RRL stars as tracers is the convolution of the true MDF of the old stellar populations, to which the RRL stars belong, and the RRL production efficiency function (RPEF) of those populations. The RPEF depends on several factors, including metallicity, age, and also helium abundance, mass-loss, etc. In general, one expects maximum RRL production for that combination of parameters that produces an even horizontal branch (HB) morphology, meaning with similar numbers of red and blue HB stars \citep[e.g.,][]{1992AJ....104.1780L,1993AJ....106.1858C}. 

It is also important to emphasize that the RPEF may well be different in the halo, the bulge and the disk, depending on how their constituent stars are distributed in terms of the aforementioned physical parameters. To achieve the same RPEF, older stellar ages are required for a metal-rich population, and younger ages for a metal-poor population, at least in the absence of systematic differences in mass loss efficiency and/or helium abundance \citep{2015pust.book.....C}. Comparing the numbers of observed local RRLs to their progenitors, \cite{1995AJ....110.2312L} estimated the RPEF for various Galactic components. Taking the halo as reference, he found the RPEF is reduced by a factor of 1/40 in the metal-rich thick disk ($[{\rm Fe/H}] > -1$), by a factor of 1/25 in the metal-poor thick disk ($[{\rm Fe/H]} < -1$), and by 1/800 in the thin disk. Therefore, we can expect the MDFs observed for the RRLs to be always skewed toward lower [Fe/H] for a generally metal-rich stellar population. As an illustration, one can consider a certain population dominated by $\sim$12~Gyr-old stars with relatively high metallicities of $[{\rm Fe/H}]\sim-0.5$, and having a long low-metallicity tail. The MDF of the resulting RRL population will not peak at $-0.5$~dex (since those give rise mostly to RC stars), but rather at $[{\rm Fe/H}] < -1$. Following the same considerations, it is clear that the ARGOS MDF, being based on mostly RC stars, also represents a `biased' version of the underlying populations' true MDF.

If the HB morphology behaves similarly in the field and in globular clusters, then we expect that the RPEF at very low metallicities is governed by the nonmonotonic nature of the HB morphology observed in halo globular clusters \citep[][and references therein]{2015pust.book.....C}. Clusters with the bluest HB types are not those at the extreme metal-poor end, but rather at $[{\rm Fe/H]} \sim -1.8$. This leads to a paucity of RRL-rich clusters at around that metallicity, since stars in clusters with very blue HBs can only become RRLs at advanced evolutionary stages of HB evolution, where they rapidly cross the instability strip, hence the probability of their observation diminishes. In contrast, several halo clusters with $[{\rm Fe/H}] < -2$ produced sizable numbers of RRL, (e.g., NGC\,7078, NGC\,4590, and NGC\,5053). It is not completely clear what causes the observed dependency of HB morphology on [Fe/H] (and thus the nonmonotonic RPEF) at low metallicities --- for a discussion of the possible physical causes, see, e.g., \cite{2008ApJ...687L..21D} and \citet[and references therein]{2016ApJ...827....2V}.

Despite the aforementioned qualitative observational and theoretical constraints on the RPEF, the current state of the art does not enable us to calibrate out its effects from the observed metallicity distribution (i.e., to deconvolve the RPEF from the RRL MDF), not just because of the lack of its quantitative description as a function of the various physical parameters, but also due to the unknowns about the distribution of those parameters in the various Galactic components. Therefore, this study does not aim to uncover the MDF of the underlying stellar populations from that of the RRL sample observed toward the southern disk, even though the primary difference between them must be due to age differences. In the future, a well-characterized RPEF will eventually allow us to quantitatively attribute the observed differences between the ARGOS and RRL MDFs to the chemical enrichment history of the Milky Way.

In Sect.~\ref{subsec:mdf}, we concluded that the MDF of the RRLs determined by our analysis is structurally similar to the MDF traced by RC stars. Our implicit assumption behind this comparison was that the composite RPEF (and similarly, the production efficiency function of RC stars) behind the observed RRL sample is monomodal, and thus its convolution with the underlying real stellar MDF cannot induce multi-modality in the observed tracer MDF. Although we previously discussed that the RPEF that governed the formation of our RRL sample is suspected to have a local minimum at $[{\rm Fe/H]} \sim -1.8$ (assuming the underlying HB population with which field RRL stars are associated behaves similarly to the trends seen among Galactic globular clusters), we argue that this cannot influence the observed means of the GMM components, since one of those lies very close to that metallicity. Therefore, under the above assumption, the observed MDF cannot be dominated by the RPEF, as far as the positions of its modes are concerned. At the same time, we must consider that if the RPEF is sharply peaked, it can induce one extra mode in the observed MDF.

\subsection{The possible origins of the observed RR~Lyrae stars}\label{subsec:origins}

Various considerations lead us to expect numerous RRLs in our sample to be part of disk populations, i.e., physically originate from the Galactic disk. Based on the observed properties of the stellar populations in the solar neighborhood, the thick disk covers an age range of 13.5--8.5 Gyr and has a metal-poor tail in its abundance distribution \citep{2015A&A...579A...5H}, so it would be surprising if the thick disk did {\em not} contain RRLs and the observed sample were instead entirely made up by interlopers from the halo. This is consistent with the existence of globular clusters on orbits confined to the Galactic disk, and abundant in RRLs (e.g., NGC\,6121 and NGC\,6626, see \citealt{1993AJ....105..168C}; and NGC\,6266, \citealt{2003AJ....125.1373D}). It might be possible that some such clusters may have long ago dissolved into the disk field, due to tidal effects, also contributing to the observed RRL population. 

The structural similarity between the RRL and the ARGOS (RC) MDFs also holds clues about the origins of the objects. It seems clear that the absence of the ARGOS component A from our data, i.e., the component consistent with a population originating from the metal-rich thin disk \citep{2013MNRAS.430..836N}, is the result of its suspected younger age \citep[e.g.][]{2013A&A...560A.109H} combined with the increasing suppression of the RPEF at high metallicities. The consistency between components B, C, and D of the ARGOS MDFs and components v1, v2, and v4 in the RRL MDF, respectively, suggests a common origin, i.e., the old thin disk and the thick disk \citep{2013MNRAS.430..836N,2013A&A...560A.109H}. The metallicity dependence of their Galactocentric distance distributions, equivalent to a radial metallicity gradient along the plane (see Sect.~\ref{subsec:spatial_mdf}), substantiates their suggested equivalency with populations emerging from the inside-out formation of the disk, as proposed by \cite{2012MNRAS.426..690B} and confirmed by \cite{2012ApJ...753..148B}. The increased concentration of components v1 and v2 toward the Galactic plane, which could be a result of their smaller scale height compared to lower-metallicity stars, is in line with this hypothesis. Also, component v4 has a mean metallicity consistent with the metal-poor extreme of the thick disk stars in the solar neighborhood \citep{2013A&A...560A.109H}, which were suggested to form a distinct population based on kinematical properties by, e.g., \cite{2010ApJ...712..692C}. The absence of an ARGOS MDF mode consistent with our component v5 is puzzling, but we note that the most metal-poor ARGOS components contain tiny amounts of stars, and therefore have very low signal-to-noise ratio. Finally, given its very low mean metallicity, stars in component v6 probably originate from interloping stellar populations, and do not physically belong to the disk. We tentatively interpret them as halo stars crossing the Galactic plane on their orbits.

The presence and dominance of component v3 (and its equivalents o3+o4) with a mean metallicity of approximately $-0.9$\,dex in the VVV disk RRL MDF, together with its complete absence from the ARGOS MDF, seem intriguing. We propose two hypotheses for the origin of this component: it may or may not be the consequence of an underlying separate stellar population. In the former case, it would result from an RPEF (see Sect.~\ref{subsec:rpef}) that is very sharply peaked at ${\rm [Fe/H]}\simeq-1$. Since the observed RRL MDF is a convolution of the parent population's MDF and the RPEF, an additional mode can appear in the former due to the functional form of the latter (assuming that the RPEF has only one significant extremum). A detailed quantitative characterization of the RPEF by means of population synthesis would be required to properly assess the possibility of this scenario. Alternatively, component v3 may arise from an underlying stellar population. In this second hypothesis, its absence from the ARGOS MDF must then be the result of the physical underrepresentation of this population in the ARGOS stellar sample (which is dominated by RC stars). This could be the case if the stars that constitute component v3 originate from a stellar population with a rather different star formation rate and age-metallicity distribution compared to the disk, resulting in a weak RC, or its complete absence in case of purely old stellar ages \citep[see, e.g.,][]{2016ARA&A..54...95G}. 

Concerning the possibility of component v3 corresponding to a separate stellar population, an important feature is the dependence of its weight on the Galactocentric distance. Let us assume that the RRL MDF components v1, v2, and v4 are indeed the equivalents of components B, C, and D in the ARGOS MDF. \cite{2013MNRAS.430..836N} clearly showed that their MDF decompositions consistently present the same components (with varying weights) when various subsamples of bulge and disk stars are concerned. This important property of the MDF has been attributed to the formation of the bulge via the dynamical instabilities of the surrounding disk \citep{2014A&A...567A.122D}. During this process, the stellar populations of the disk flare up into a peanut-shaped configuration observed in the central region of the Milky Way, while preserving the multi-modality of their MDF. The efficiency of this spatial redistribution depends on the initial dynamical conditions in the sense that a dynamically hotter population will be less perturbed, and this will manifest itself as the spatial dependence in the weights of the observed MDF components. Following this bulge formation scenario, the disk RRL populations that correspond to components v1, v2, and v4 must have also witnessed the same perturbations, which must have happened at a rather early epoch, given the fact that the RRLs are more than 10~Gyr old. As a result of this, the bulge RRL MDF must contain contributions from these populations, which are observed as the GMM components o1, o2, and o5, identified in the weak tails of the distribution. Their very small weights in the MDF may arise from the fact that the population pertaining to the GMM component o4 at ${\rm [Fe/H]}\simeq-1$ suppresses the other components, on account of its large relative fraction in the bulge sample, which gradually diminishes further away from the Galactic Center. Such a population of stars could possibly originate from a primordial gravitational collapse or hierarchical merging of protogalactic fragments, commonly referred to as an ``inner halo'' and a ``classical spheroid'' \citep[see, e.g.,][]{2010A&A...519A..77B}. In the bulge, observational evidence for the existence of a spheroidal component has been provided by the structural and kinematical properties of its constituent RRL stars, namely that they are systematically different from those of the red clump giants \citep{2013ApJ...776L..19D,2016ApJ...821L..25K}. Stars contributing to component 3 in the RRL MDF observed toward the disk might be the weak outer part of this spheroid, overlapping the inner disk populations.

Further evidence for the origin of the various components of the RRL sample presented here could be provided by time-series photometric studies of the disk covering a more extended range in Galactic latitude. Those RRLs that belong to the thick disk should show the same Galactic spatial distribution as the red giants that were used to trace the thick disk. Conversely, stars belonging to the spheroid would not show a higher density along the disk plane. Our current sample is constrained well below the scale height of the thick disk, but the OGLE-IV survey and the ongoing VVVX survey have wider footprints, so the RRL census from those will provide additional key information about the origin of the RRLs observed toward the disk. Kinematical information will also be crucial for confirming the provenance of the RRLs, since the orbital motions of stars that physically belong to different Galactic components, such as disk, bulge, and halo, are systematically different. The VVV survey yielded a database of proper motions \citep[VIRAC,][]{2018MNRAS.474.1826S}, which will provide useful kinematic information for a considerable fraction of our sample. The astrometry to be provided by future Gaia data releases will be complementary, and parallax measurements for objects lying at the near side of the disk will enable us to decrease the error in the distance distribution. We will extend our analysis of RRLs by incorporating kinematical data in a forthcoming paper.

\section{Summary} \label{sec:summary}

By performing a deep census of RRL stars along the southern Galactic disk, we uncovered invaluable tracers of the Milky Way's stellar fossil, which provide important constraints on our Galaxy's early evolution history. The strong constraint on the ages of RRLs helps us to better understand the formation and chemical enrichment history of the Milky Way's disk populations. Although the current lack of a quantitative understanding of the RRL production efficiency function hinders a direct measurement of the MDF of the underlying stellar populations, the comparison of the observed [Fe/H] distribution to independent abundance measurements of clump giants by the ARGOS survey revealed significant similarities, which hints at multiple RRL populations toward the disk, probably originating from different Galactic components.

In contrast to the intermediate-age sample from ARGOS, their most metal-rich and prominent component at a mean metallicity of [Fe/H]$\simeq$+0.1~dex, which they attribute to the metal-rich disk (and its heated progeny in the bulge), is essentially invisible in the RRL population. The populations indicated by the two metal-rich modes in the RRL metallicity distribution at $[{\rm Fe/H}]\simeq-0.3$ and $[{\rm Fe/H}]\simeq-0.6$ were presumably associated with the metal-poor thin disk and the thick disk, respectively, while the metal-poor mode at $[{\rm Fe/H}] \simeq -1.2$ could possibly be the equivalent of a distinct metal-poor thick disk population.
Similar to the bulge, stars with metallicities of $[{\rm Fe/H}] \simeq -1$ dominate the MDF at all Galactic distances in our disk sample, which may belong to a classical spheroid, while the small number of metal-poor ([${\rm Fe/H}]\lesssim-1.4$) RRL might be halo stars on orbits currently crossing the disk. New kinematical data on the objects from the VVVX and the Gaia surveys, the extension of the present census by the OGLE-IV survey, and accurate spectroscopic abundance measurements from upcoming spectroscopic surveys will enable us to scrutinize our current, tentative interpretation of the RRL stars. Ultimately, the RRL stars presented here will enable us to put key constraints on the evolution models of the Milky Way's bulge and disk and better understand their various aspects, such as the chemical enrichment and structural evolution through dynamical instabilities. 

\acknowledgments

I.D. and E.K.G. were supported by Sonderforschungsbereich SFB 881 ``The Milky Way System'' (subproject A3) of the German Research Foundation (DFG).
G.H. acknowledges support from the Graduate Student Exchange Fellowship Program between the Institute of Astrophysics of the Pontificia Universidad Cat\'olica de Chile and the Zentrum f\"ur Astronomie der Universit\"at Heidelberg, funded by the Heidelberg Center in Santiago de Chile and the Deutscher Akademischer Austauschdienst (DAAD) and by CONICYT-PCHA/Doctorado Nacional grant 2014-63140099.
G.H., M.C., F.E., S.E., and A.J. also acknowledge support by the Ministry for the Economy, Development, and Tourism's Programa Iniciativa Milenio through grant IC120009 and by Proyecto Basal PFB-06/2007.
G.H. and M.C. acknowledge additional support by FONDECYT grant \#1171273 and by CONICYT's PCI program through grant DPI20140066.
F.E. acknowledges support from CONICYT-PCHA (Doctorado Nacional 2014-21140566).
M.C. gratefully acknowledges additional support by the DAAD and the DFG.
Post-processing and analysis of data were performed on the Milky Way supercomputer, which is funded by the Deutsche Forschungsgemeinschaft (DFG) through the Collaborative Research Center (SFB 881) ``The Milky Way System'' (subproject Z2).

\vspace{5mm}
\facility{ESO:VISTA}
\software{STIL \citep{2006ASPC..351..666T}, 
                scikit-learn \citep{2011JMLR....12.2825}
          }


\begin{thebibliography}{}

\bibitem[Akaike(1974)]{1974IEEE..19....6} Akaike,~H.\ 1974, IEEE Transactions on Automatic Control, 19, 6

\bibitem[Babusiaux et al.(2010)]{2010A&A...519A..77B} Babusiaux, C., G{\'o}mez, A., Hill, V., et al.\ 2010, \aap, 519, A77 

\bibitem[Beaton et al.(2016)]{2016ApJ...832..210B} Beaton, R.~L., Freedman, W.~L., Madore, B.~F., et al.\ 2016, \apj, 832, 210 

\bibitem[Bovy et al.(2012)]{2012ApJ...753..148B} Bovy, J., Rix, H.-W., Liu, C., et al.\ 2012, \apj, 753, 148 

\bibitem[Braga et al.(2015)]{2015ApJ...799..165B} Braga, V.~F., Dall'Ora, M., Bono, G., et al.\ 2015, \apj, 799, 165 

\bibitem[Braga et al.(2016)]{2016AJ....152..170B} Braga, V.~F., Stetson, P.~B., Bono, G., et al.\ 2016, \aj, 152, 170 

\bibitem[Brook et al.(2012)]{2012MNRAS.426..690B} Brook, C.~B., Stinson, G.~S., Gibson, B.~K., et al.\ 2012, \mnras, 426, 690 

\bibitem[Brown et al.(2010)]{2010ApJ...725L..19B} Brown, T.~M., Sahu, K., Anderson, J., et al.\ 2010, \apjl, 725, L19

\bibitem[Carollo et al.(2010)]{2010ApJ...712..692C} Carollo, D., Beers, T.~C., Chiba, M., et al.\ 2010, \apj, 712, 692 

\bibitem[Carpenter (2001)]{2001AJ....121.2851C} Carpenter, J.~M.\ 2001, \aj, 121, 2851

\bibitem[Carretta et al.(2009)]{2009A&A...508..695C} Carretta, E., Bragaglia, A., Gratton, R., D'Orazi, V., \& Lucatello, S.\ 2009, \aap, 508, 695 

\bibitem[Catelan \& de Freitas Pacheco(1993)]{1993AJ....106.1858C} Catelan, M., \& de Freitas Pacheco, J.~A.\ 1993, \aj, 106, 1858 

\bibitem[Catelan et al.(2004)]{2004ApJS..154..633C} Catelan, M., Pritzl, B.~J., \& Smith, H.~A.\ 2004, \apjs, 154, 633

\bibitem[Catelan(2009)]{2009IAUS..258..209C} Catelan, M.\ 2009, The Ages of Stars, 258, 209 

\bibitem[Catelan \& Smith(2015)]{2015pust.book.....C} Catelan, M., \& Smith, H.~A.\ 2015, Pulsating Stars (Wiley-VCH)

\bibitem[Cudworth \& Hanson(1993)]{1993AJ....105..168C} Cudworth, K.~M., \& Hanson, R.~B.\ 1993, \aj, 105, 168 

\bibitem[Dall'Ora et al.(2004)]{2004ApJ...610..269D} Dall'Ora, M., Storm, J., Bono, G., et al.\ 2004, \apj, 610, 269 

\bibitem[Di Matteo et al.(2014)]{2014A&A...567A.122D} Di Matteo, P., Haywood, M., G{\'o}mez, A., et al.\ 2014, \aap, 567, A122 
%

\bibitem[de Grijs \& Bono(2016)]{2016ApJS..227....5D} de Grijs, R., \& Bono, G.\ 2016, \apjs, 227, 5 

\bibitem[D{\'e}k{\'a}ny et al.(2013)]{2013ApJ...776L..19D} D{\'e}k{\'a}ny, I., Minniti, D., Catelan, M., et al.\ 2013, \apjl, 776, L19 

\bibitem[Dinescu et al.(2003)]{2003AJ....125.1373D} Dinescu, D.~I., Girard, T.~M., van Altena, W.~F., \& L{\'o}pez, C.~E.\ 2003, \aj, 125, 1373 

\bibitem[Dotter(2008)]{2008ApJ...687L..21D} Dotter, A.\ 2008, \apjl, 687, L21 

\bibitem[Drake et al.(2013)]{2013ApJ...763...32D} Drake, A.~J., Catelan, M., Djorgovski, S.~G., et al.\ 2013, \apj, 763, 32 

\bibitem[Elorrieta et al.(2016)]{2016A&A...595A..82E} Elorrieta, F., Eyheramendy, S., Jord{\'a}n, A., et al.\ 2016, \aap, 595, A82 

\bibitem[Emerson et al.(2004)]{2004SPIE.5493..401E} Emerson, J.~P., Irwin, M.~J., Lewis, J., et al.\ 2004, \procspie, 5493, 401 

\bibitem[Ferguson et al.(2005)]{2005ApJ...623..585F} Ferguson, J.~W., Alexander, D.~R., Allard, F., et al.\ 2005, \apj, 623, 585 

\bibitem[Fiorentino et al.(2017)]{2017A&A...599A.125F} Fiorentino, G., Monelli, M., Stetson, P.~B., et al.\ 2017, \aap, 599, A125 

\bibitem[Freeman et al.(2013)]{2013MNRAS.428.3660F} Freeman, K., Ness, M., Wylie-de-Boer, E., et al.\ 2013, \mnras, 428, 3660 

\bibitem[Gillessen et al.(2009)]{2009ApJ...692.1075G} Gillessen, S., Eisenhauer, F., Trippe, S., et al.\ 2009, \apj, 692, 1075 

\bibitem[Girardi \& Salaris(2001)]{2001MNRAS.323..109G} Girardi, L., \& Salaris, M.\ 2001, \mnras, 323, 109 

\bibitem[Girardi(2016)]{2016ARA&A..54...95G} Girardi, L.\ 2016, \araa, 54, 95 

\bibitem[Glatt et al.(2008)]{2008AJ....135.1106G} Glatt, K., Gallagher, J.~S., III, Grebel, E.~K., et al.\ 2008, \aj, 135, 1106 

\bibitem[Grevesse \& Sauval(1998)]{1998SSRv...85..161G} Grevesse, N., \& Sauval, A.~J.\ 1998, \ssr, 85, 161 

\bibitem[Hajdu et al.(2018)]{hajdu} Hajdu, G., D\'ek\'any, I., Catelan, M., Grebel, E.~K., Jurcsik, J.\ 2018, \apj, in press

\bibitem[Hansen et al.(2011)]{2011ApJ...743L...1H} Hansen, T., Andersen, J., Nordstr{\"o}m, B., Buchhave, L.~A., \& Beers, T.~C.\ 2011, \apjl, 743, L1 

\bibitem[Hayden et al.(2015)]{2015ApJ...808..132H} Hayden, M.~R., Bovy, J., Holtzman, J.~A., et al.\ 2015, \apj, 808, 132 

\bibitem[Haywood et al.(2013)]{2013A&A...560A.109H} Haywood, M., Di Matteo, P., Lehnert, M.~D., Katz, D., \& G{\'o}mez, A.\ 2013, \aap, 560, A109 

\bibitem[Haywood et al.(2015)]{2015A&A...579A...5H} Haywood, M., Di Matteo, P., Snaith, O., \& Lehnert, M.~D.\ 2015, \aap, 579, A5 

\bibitem[Irwin et al.(2004)]{2004SPIE.5493..411I} Irwin, M.~J., Lewis, J., Hodgkin, S., et al.\ 2004, \procspie, 5493, 411 

\bibitem[Ivezi{\'c} et al.(2014)]{2014sdmm.book.....I} Ivezi{\'c}, {\v Z}., Connelly, A.~J., VanderPlas, J.~T., \& Gray, A.\ 2014, Statistics, Data Mining, and Machine Learning in Astronomy,~Princeton, NJ: Princeton University Press

\bibitem[Jurcsik(1995)]{1995AcA....45..653J} Jurcsik, J.\ 1995, \actaa, 45, 653 
%

\bibitem[Jurcsik \& Kov\'acs(1996)]{1996A&A...312..111J} Jurcsik, J., \& Kov\'acs, G.\ 1996, \aap, 312, 111 

\bibitem[Kov{\'a}cs \& Kupi(2007)]{2007A&A...462.1007K} Kov{\'a}cs, G., \& Kupi, G.\ 2007, \aap, 462, 1007 

\bibitem[Kov\'acs \& Zsoldos(1995)]{1995A&A...293L..57K} Kov\'acs, G., \& Zsoldos, E.\ 1995, \aap, 293, L57

\bibitem[Kunder et al.(2016)]{2016ApJ...821L..25K} Kunder, A., Rich, R.~M., Koch, A., et al.\ 2016, \apjl, 821, L25 

\bibitem[Layden(1994)]{1994AJ....108.1016L} Layden, A.~C.\ 1994, \aj, 108, 1016 

\bibitem[Layden(1995)]{1995AJ....110.2312L} Layden, A.~C.\ 1995, \aj, 110, 2312 


\bibitem[Lee(1992)]{1992AJ....104.1780L} Lee, Y.-W.\ 1992, \aj, 104, 1780 

\bibitem[Liddle(2007)]{2007MNRAS.377L..74L} Liddle, A.~R.\ 2007, \mnras, 377, L74 

\bibitem[Majaess et al.(2016)]{2016A&A...593A.124M} Majaess, D., Turner, D., D{\'e}k{\'a}ny, I., Minniti, D., \& Gieren, W.\ 2016, \aap, 593, A124 

\bibitem[Marconi et al.(2015)]{2015ApJ...808...50M} Marconi, M., Coppola, G., Bono, G., et al.\ 2015, \apj, 808, 50 

\bibitem[Mart{\'{\i}}nez-V{\'a}zquez et al.(2015)]{2015MNRAS.454.1509M} Mart{\'{\i}}nez-V{\'a}zquez, C.~E., Monelli, M., Bono, G., et al.\ 2015, \mnras, 454, 1509 

\bibitem[Mart{\'{\i}}nez-V{\'a}zquez et al.(2016)]{2016MNRAS.461L..41M} Mart{\'{\i}}nez-V{\'a}zquez, C.~E., Monelli, M., Gallart, C., et al.\ 2016, \mnras, 461, L41 

\bibitem[Minniti et al.(2010)]{2010NewA...15..433M} Minniti, D., Lucas, P.~W., Emerson, J.~P., et al.\ 2010, \na, 15, 433 

\bibitem[Minniti et al.(2017)]{2017AJ....153..179M} Minniti, D., D{\'e}k{\'a}ny, I., Majaess, D., et al.\ 2017, \aj, 153, 179 

\bibitem[Minniti(2018)]{2018ASSP...51...63M} Minniti, D.\ 2018, The Vatican Observatory, Castel Gandolfo: 80th Anniversary Celebration, 51, 63 

\bibitem[Muraveva et al.(2015)]{2015ApJ...807..127M} Muraveva, T., Palmer, M., Clementini, G., et al.\ 2015, \apj, 807, 127 

\bibitem[Nataf et al.(2016)]{2016MNRAS.456.2692N} Nataf, D.~M., Gonzalez, O.~A., Casagrande, L., et al.\ 2016, \mnras, 456, 2692 

\bibitem[Navarrete et al.(2017)]{2017A&A...604A.120N} Navarrete, C., Catelan, M., Contreras Ramos, R., et al.\ 2017, \aap, 604, A120; erratum: 2017, A\&A, 606, C1

\bibitem[Ness et al.(2013)]{2013MNRAS.430..836N} Ness, M., Freeman, K., Athanassoula, E., et al.\ 2013, \mnras, 430, 836 

\bibitem[Nidever et al.(2014)]{2014ApJ...796...38N} Nidever, D.~L., Bovy, J., Bird, J.~C., et al.\ 2014, \apj, 796, 38 

\bibitem[Oosterhoff(1939)]{1939Obs....62..104O} Oosterhoff, P.~T.\ 1939, The Observatory, 62, 104 

\bibitem[Ortolani et al.(1995)]{1995Natur.377..701O} Ortolani, S., Renzini, A., Gilmozzi, R., et al.\ 1995, \nat, 377, 701 

\bibitem[Pedregosa et al.(2011)]{2011JMLR....12.2825} Pedregosa, F., Varoquaux, G., Gramfort, A., et al.\ 2011, Journal of Machine Learning Research, 12, 2825

\bibitem[Pietrukowicz et al.(2015)]{2015ApJ...811..113P} Pietrukowicz, P., Koz{\l}owski, S., Skowron, J., et al.\ 2015, \apj, 811, 113 

\bibitem[Saito et al.(2012)]{2012A&A...537A.107S} Saito, R.~K., Hempel, M., Minniti, D., et al.\ 2012, \aap, 537, A107

\bibitem[Schlafly et al.(2016)]{2016ApJ...821...78S} Schlafly, E.~F., Meisner, A.~M., Stutz, A.~M., et al.\ 2016, \apj, 821, 78 

\bibitem[Schlegel et al.(1998)]{1998ApJ...500..525S} Schlegel, D.~J., Finkbeiner, D.~P., \& Davis, M.\ 1998, \apj, 500, 525 

\bibitem[Sesar et al.(2013)]{2013AJ....146...21S} Sesar, B., Ivezi{\'c}, {\v Z}., Stuart, J.~S., et al.\ 2013, \aj, 146, 21 

\bibitem[Skrutskie et al.(2006)]{2006AJ....131.1163S} Skrutskie, M.~F., Cutri, R.~M., Stiening, R., et al.\ 2006, \aj, 131, 1163 

\bibitem[Smith et al.(2018)]{2018MNRAS.474.1826S} Smith, L.~C., Lucas, P.~W., Kurtev, R., et al.\ 2018, \mnras, 474, 1826 

\bibitem[Smolec(2005)]{2005AcA....55...59S} Smolec, R.\ 2005, \actaa, 55, 59 

\bibitem[Soszy{\'n}ski et al.(2014)]{2014AcA....64..177S} Soszy{\'n}ski, I., Udalski, A., Szyma{\'n}ski, M.~K., et al.\ 2014, \actaa, 64, 177

\bibitem[Soszy{\'n}ski et al.(2016)]{2016AcA....66..131S} Soszy{\'n}ski, I., Udalski, A., Szyma{\'n}ski, M.~K., et al.\ 2016, \actaa, 66, 131 

\bibitem[Stetson(1996)]{1996PASP..108..851S} Stetson, P.~B.\ 1996, \pasp, 108, 851 

\bibitem[Udalski et al.(2015)]{2015AcA....65....1U} Udalski, A., Szyma{\'n}ski, M.~K., \& Szyma{\'n}ski, G.\ 2015, \actaa, 65, 1 

\bibitem[Taylor(2006)]{2006ASPC..351..666T} Taylor, M.~B.\ 2006, in Astronomical Data Analysis Software and Systems XV (ed. C. Gabriel et al.), ASP Conference Series, 351, p. 666 

\bibitem[VandenBerg et al.(2016)]{2016ApJ...827....2V} VandenBerg, D.~A., Denissenkov, P.~A., \& Catelan, M.\ 2016, \apj, 827, 2 

\bibitem[von Neumann et al.(1941)]{1941AMS..12..153}von Neumann, J., Kent, R. H., Bellinson, H. R., Hart, B. I.\ 1941, The Annals of Mathematical Statistics, 12, 153

\bibitem[Zechmeister \& K{\"u}rster(2009)]{2009A&A...496..577Z} Zechmeister, M., \& K{\"u}rster, M.\ 2009, \aap, 496, 577 

\end{thebibliography}
\end{document}